\documentclass[prl,aps,twocolumn,superscriptaddress,showpacs]{revtex4-1}
\usepackage{graphicx}
\usepackage{amsmath}
\usepackage{epsfig}
\usepackage{multirow}
\usepackage{array}
\usepackage{inputenc}
\usepackage{xcolor}
\usepackage{tabularx}
\usepackage{textcomp}
\usepackage{ulem}

\begin{document}


\title{From one- to two-magnon excitations in the S = 3/2 magnet $\beta$-CaCr$_2$O$_4$}

\author{M. Songvilay}
\altaffiliation{present address: Institut N\'eel, CNRS \& Univ. Grenoble Alpes, 38000 Grenoble, France}
\affiliation{School of Physics and Astronomy, University of Edinburgh, Edinburgh EH9 3JZ, United Kingdom}

\author{S. Petit}
\affiliation{Laboratoire L\'{e}on Brillouin, CEA-CNRS UMR 12, 91191 Gif-Sur-Yvette Cedex, France}

\author{F. Damay}
\affiliation{Laboratoire L\'{e}on Brillouin, CEA-CNRS UMR 12, 91191 Gif-Sur-Yvette Cedex, France}

\author{G. Roux}
\affiliation{Universit\'e Paris-Saclay, CNRS, LPTMS, 91405 Orsay, France}

\author{N. Qureshi}
\affiliation{Institut Laue-Langevin, 6 rue Jules Horowitz, Boite postale 156, 38042 Grenoble, France}

\author{H. C. Walker}
\affiliation{ISIS Neutron and Muon Source, Rutherford Appleton Laboratory, Chilton, Didcot OX11 0QX, United Kingdom}

\author{J. A. Rodriguez-Rivera}
\affiliation{NIST Center for Neutron Research, National Institute of Standards and Technology, 100 Bureau Drive, Gaithersburg, Maryland, 20899, USA}
\affiliation{Department of Materials Science, University of Maryland, College Park, Maryland 20742, USA}

\author{B. Gao}
\affiliation{Rutgers Center for Emergent Materials and Department of Physics and Astronomy, Rutgers University, 136 Frelinghuysen Road, Piscataway, New Jersey 08854, USA}

\author{S.-W. Cheong}
\affiliation{Rutgers Center for Emergent Materials and Department of Physics and Astronomy, Rutgers University, 136 Frelinghuysen Road, Piscataway, New Jersey 08854, USA}

\author{C. Stock}
\affiliation{School of Physics and Astronomy, University of Edinburgh, Edinburgh EH9 3JZ, United Kingdom}

\date{\today}

\begin{abstract}

We apply neutron spectroscopy to measure the magnetic dynamics in the $S$~=~3/2 magnet $\beta$-CaCr$_2$O$_4$ (T$_{N}$=21 K).  The low-energy fluctuations, in the ordered state, resemble large-$S$ linear spin-waves from the incommensurate ground state.   However, at higher energy transfers, these semiclassical and harmonic dynamics are replaced  by an energy and momentum broadened continuum of excitations.  Applying kinematic constraints required for energy and momentum conservation, sum rules of neutron scattering, and comparison against exact diagonalization calculations, we show that the dynamics at high-energy transfers resemble low-$S$ one-dimensional quantum fluctuations.  $\beta$-CaCr$_2$O$_4$ represents an example of a magnet at the border between classical N\'eel and quantum phases, displaying dual characteristics.

\end{abstract}

\pacs{}

\maketitle


Quantum fluctuations originate from the uncertainty inherent to non commuting observables and appear through the neutron scattering cross section in one-dimensional $S$~=~1/2 Heisenberg spin chains~\cite{Bethe1931,Cloizeaux1962,Muller1981}.  The excitations in these magnets are spinons and manifest as a momentum and energy broadened continuum in the neutron response  that correspond to domain boundaries from pairs of spins which disrupt the N\'eel order~\cite{Nagler1991,Tennant1993, Lake2013,Mourigal2013,Stone2003,Kenzelmann2004,Enderle10:104}.  This contrasts to the case of classical large-$S$ spin waves that are long-lived harmonic precessions around a spatially ordered magnetic ground state~\cite{Hutchings1972}.  The neutron response in this latter case is characterized by well-defined excitation dispersion.

The quantum and classical cases represent two extremes that are treated differently with the Bethe ansatz applied to low-$S$ cases~\cite{Bethe1931,Cloizeaux1962} and semiclassical quantization based on transverse motions of the spin around an ordered moment direction applied to large-$S$.  Quantum fluctuations are enhanced in low-spin and low-dimensional magnets and such longitudinal continua of excitations have been extensively studied in $S$~=~1/2 chains \cite{Endoh1974,Heilmann1978,Schulz1996, Essler1997, Lake2005, Lake2010, Zheludev2002, Zheludev2002bis, Zheludev2003}. While classical spin-waves dominate the cross section of large-$S$ magnets, weak quantum corrections exist in large-$S$~=~5/2 low dimensional magnets \cite{Heilmann1981,Huberman2005, Songvilay2018}. We investigate the $S$~=~3/2, $\beta$-CaCr$_2$O$_4$ magnet, where both of these extremes are present with classical linear spin-waves breaking down into quantum fluctuations displaying a dual quantum/classical character.

\begin{figure}[t]
 \includegraphics[scale=0.35]{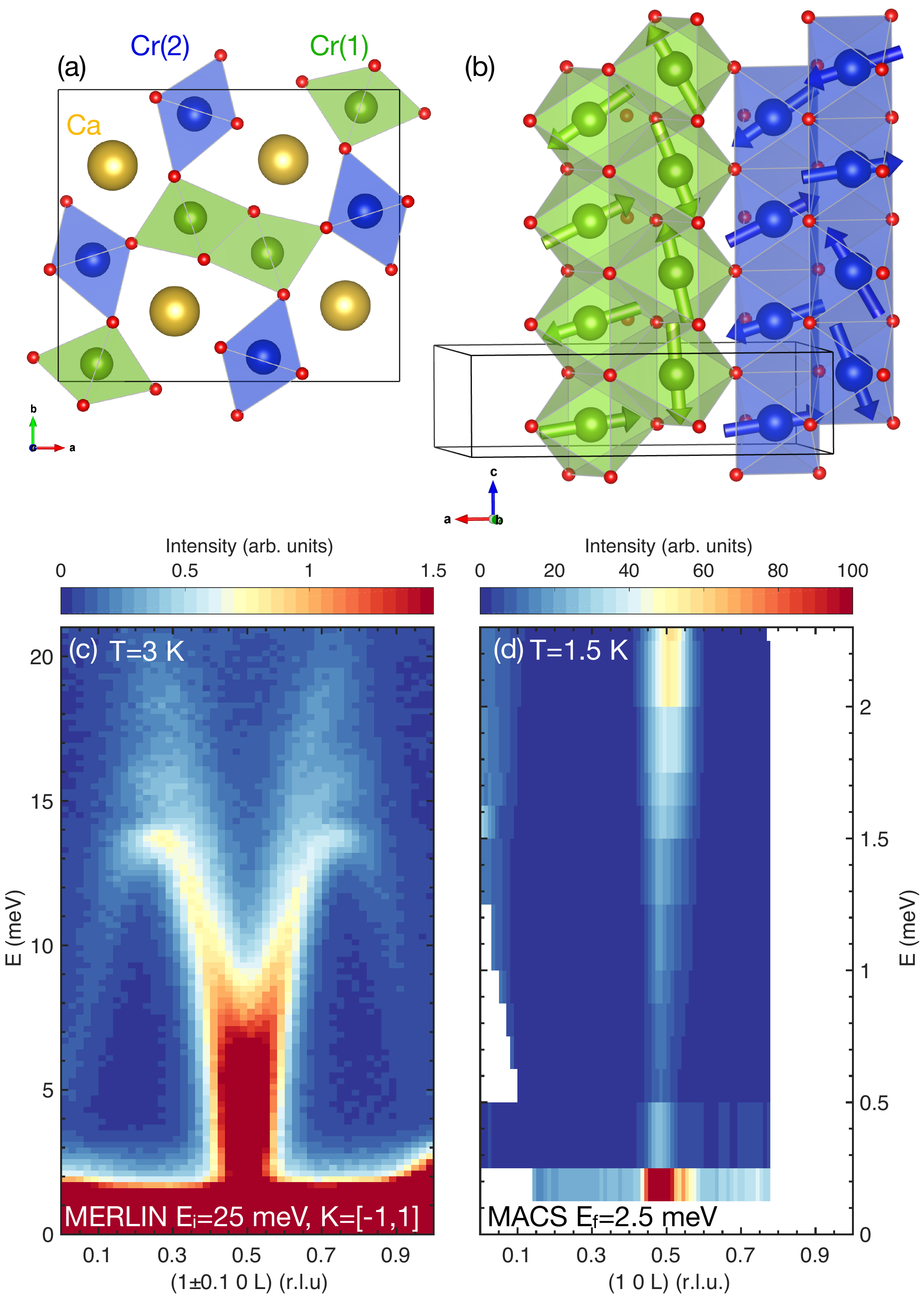}
 \caption{ \label{fig:dispersion} 
 $(a)$ Crystallographic structure of $\beta$-CaCr$_{2}$O$_{4}$.  View of the network of CrO$_{6}$ octahedra in the $a-b$ plane.  $(b)$ View of the zig-zag chains of Cr$^{3+}$ in the $a-c$ plane and the associated incommensurate magnetic structure. $(c)$ Colormap of the magnetic excitations along (1 0 L) measured on MERLIN. $(d)$ High resolution neutron inelastic data taken on MACS.}
\end{figure}

$\beta$-CaCr$_2$O$_4$ is orthorhombic (space group 62 $Pbnm$ with $a$=10.61, $b$=9.09, $c$=2.94 \AA), isostructural with CaFe$_2$O$_4$~\cite{Decker1957,Hill1956,Corliss1967}.  The CrO$_6$ octahedra (Figs. \ref{fig:dispersion} $(a,b)$) form edge sharing chains along the $c$-axis with the Cr positions in any adjacent chain translated along $c$ by (0~0~1/2), so that inter-chain coupling, either along the $a$ or $b$-axis, has an anisotropic triangular arrangement (Fig. \ref{fig:dispersion} $(a,b)$). Spatially long-ranged incommensurate cycloidal magnetic ordering with a propagation vector \textbf{k}~=~(0,~0,~$\sim$~0.477) occurs below T$_{N}$=21 K~\cite{Damay2010,Damay2011}. The refined Cr$^{3+}$ ($S$~=~3/2, $L$=0) ordered magnetic moment (2.88 $\mu_B$) is reduced compared to $gS$~=~3~$\mu_B$ ($g$~=~2 is the Land\'e factor).

We apply neutron spectroscopy to investigate the magnetic excitations, using a mirror furnace synthesized single crystal of $\beta$-CaCr$_2$O$_4$. Measurements were performed on the MERLIN (ISIS, UK) \cite{MERLIN},  2T1 (Laboratoire L\'eon Brillouin-Orph\'ee, France) and MACS (NIST, US) spectrometers. Further details are supplied in the Supplemental Material.

\begin{figure}[t]
 \includegraphics[scale=0.34]{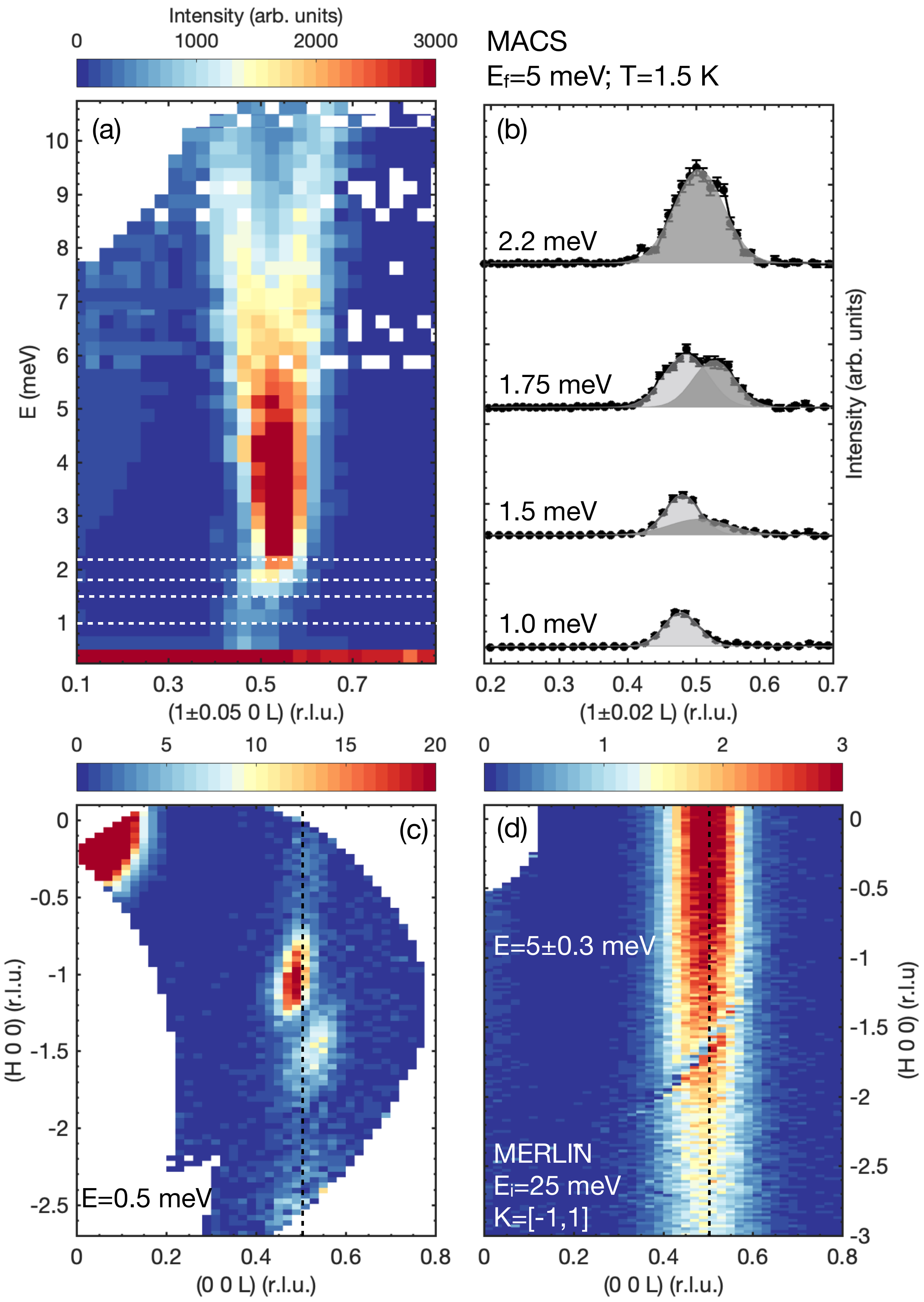}
 \caption{ \label{fig:macs_lowT} $(a)$ Low-energy neutron inelastic data (MACS, E$_{f}$=5 meV, T=1.5 K) with $(b)$ constant energy cuts.  Gray areas indicate the gaussian fits.  $(c)$ Constant energy slices (MACS).  $(d)$ Similar slice at 5 meV from MERLIN (E$_{i}$=25 meV).}
\end{figure} 

We first discuss the magnetic dynamics along the $c$-axis, along which strongly bonded CrO$_6$ octahedra form chains (Fig. \ref{fig:dispersion} $(a,b)$) and are hence expected to display strong magnetic coupling.  This is reflected in isostructural CaFe$_{2}$O$_{4}$~\cite{Stock2016,Stock2017} which shows a well-defined single-magnon branch gapped due to anisotropy.  The full magnetic excitation spectrum in the N\'eel state (T~=~3 K) of $\beta$-CaCr$_{2}$O$_{4}$ is illustrated in Figures \ref{fig:dispersion} $(c)$ and $(d)$ extending up to $\sim$ 15 meV.  Fig. \ref{fig:dispersion} $(c)$ illustrates a strongly dispersive excitation that emanates from the commensurate L=0.5 position extending to $\sim$ 13 meV indicative of strong magnetic coupling along $c$.  However, there are two features that distinguish CaCr$_{2}$O$_{4}$ from the isostructural $S$=5/2, CaFe$_{2}$O$_{4}$.  First, as displayed in Fig. \ref{fig:dispersion} $(c)$, two weaker and energy broadened excitations are present above $\sim$ 15 meV.  Second, as shown in Fig. \ref{fig:dispersion} $(d)$, the excitations cross over at lower energies to become incommensurate and gapless within resolution.  We address both these points below.

Figure \ref{fig:macs_lowT} $(a)$ shows the low-energy magnetic scattering in the (H 0 L) plane. Fig. \ref{fig:macs_lowT} $(a)$ illustrates a constant momentum slice from MACS showing a $\sim$ 3 meV gapped commensurate response and a lower energy incommensurate response.  This is highlighted by a series of constant energy cuts (indicated by the dashed lines) and plotted in Fig. \ref{fig:macs_lowT} $(b)$.  At 1.0 meV, an incommensurate peak is found centered at (-1 0 0.47).  With increasing energy transfer, the peak emanating from the incommensurate magnetic position gradually shifts to the antiferromagnetic position L~=~0.5 at E~=~2.2~meV.    The momentum dependence in the (H 0 L) plane is further investigated in constant energy slices in Figure \ref{fig:macs_lowT}  $(c,d)$.   Fig. \ref{fig:macs_lowT} $(c)$ (E~=~0.5~meV) displays a peak well defined in momentum along (H~0~0) and (0~0~L) and centered around the incommensurate (-1 0 0.47) position. A weaker peak exists near (-1.5 0 $\sim$0.55) and is also present at E~=~0 meV.   While this peak disappears with increased temperature, suggestive of a magnetic origin, a neutron polarization analysis using {\sc cryopad} finds it \textit{not} to be fully magnetic as discussed in the Supplementary Information.  Hence, this feature will not be discussed further.  As shown in Figure \ref{fig:macs_lowT}  $(d)$, with increasing energy transfer, the peak emanating from the incommensurate magnetic position broadens along H, while remaining well defined along L.  This indicates a loss of dynamic spin correlations along the $a$-axis while strong correlations along the $c$-axis remain where a bonded Cr$^{3+}$ network exists (Fig. \ref{fig:dispersion}).  Constant energy scans (Supplementary Information) show that the spins are also uncorrelated along the $b$-axis at energy transfers above at least $\sim$ 1 meV consistent with results on isostructural CaFe$_{2}$O$_{4}$~\cite{Stock2016}.  The low energy spin dynamics below $\sim$ 2 meV are indicative of two dimensional spin waves while the loss of correlations along H at higher energy transfers illustrates one dimensional dynamic correlations.

\begin{figure}
 \includegraphics[scale=0.30]{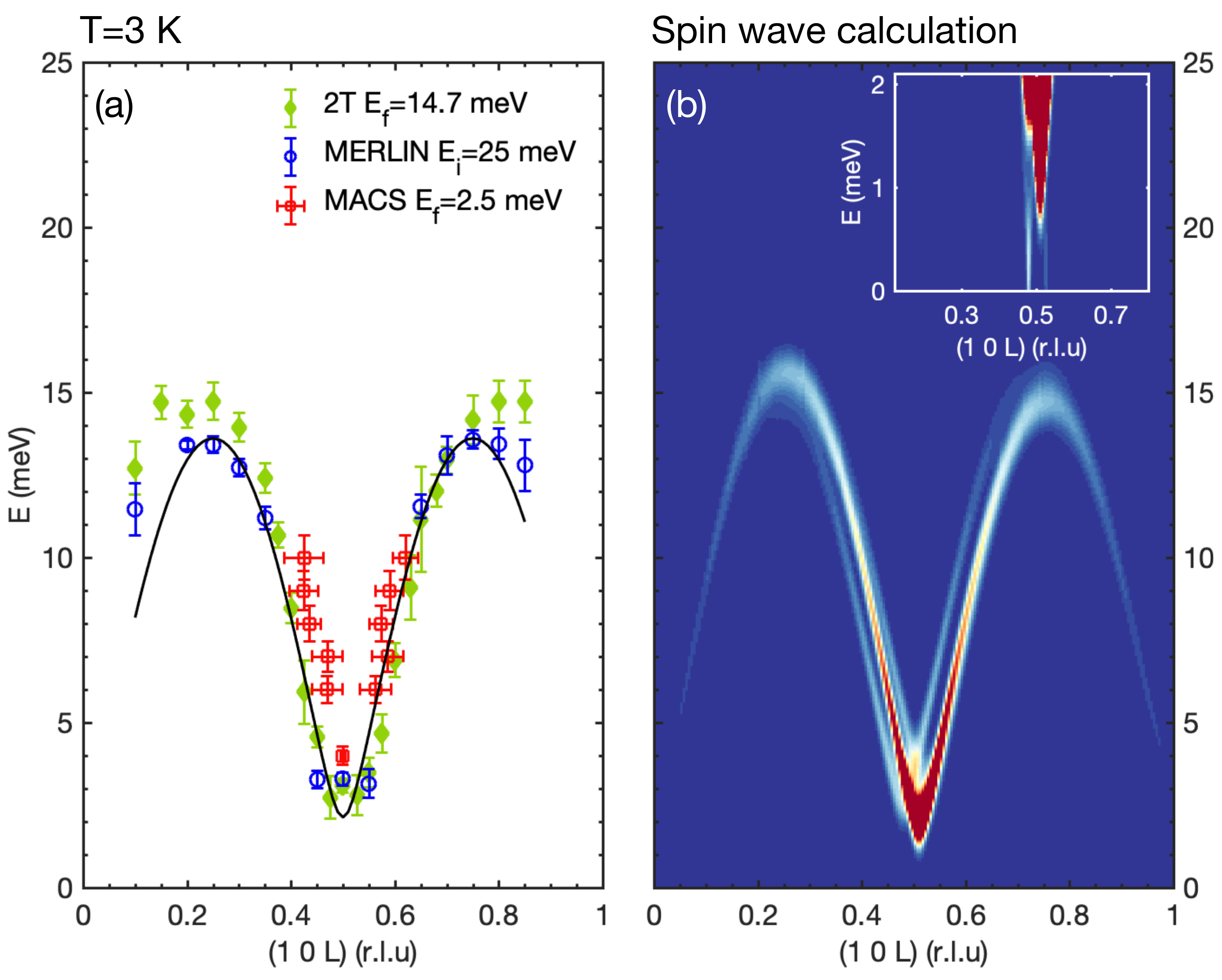}
 \caption{ \label{fig:macs_vsE} Comparison between the measured magnetic dispersion $(a)$ and calculated $(b)$ using linear spin-wave theory.}
\end{figure}

To understand both the gapped commensurate dispersive mode and the low-energy gapless incommensurate response, we compare the measured dispersion to linear spin-wave theory in Figure \ref{fig:macs_vsE}.  Fig. \ref{fig:macs_vsE} $(a)$ illustrates the gapped commensurate response compared to linear spin-wave theory based on the refined cycloidal magnetic ground state in Fig. \ref{fig:macs_vsE} $(b)$. The calculated dispersion is in agreement with Figure \ref{fig:dispersion} $(c)$ reproducing the dispersing mode from (1 0 0.5), but the inset illustrates the prediction of a low-energy gapless mode emanating from the incommensurate magnetic Bragg position.  The calculated gapped mode with a dispersion minimum at the commensurate antiferromagnetic (1~0~0.5) position corresponds to out-of-plane spin fluctuations. Goldstone modes originating from the incommensurate magnetic ordering wavevector are observed at energies below $\sim$ 2 meV (Figure \ref{fig:macs_lowT}), as calculated in the inset of Figure \ref{fig:macs_vsE} $(b)$. This model, describes most of the high-resolution data measured on MACS (Fig. \ref{fig:dispersion} $(d)$), although some discrepancies remain between the model and the reported magnetic groundstate, as detailed in the Supplemental Information. This classical spin-wave model is based on transverse fluctuations from a large-$S$ ground state, and these fluctuations can be interpreted quantum mechanically as single magnon excitations~\cite{Holstein1940,Dyson1956}.

However, as noted above, Figure  \ref{fig:dispersion} $(c)$ displays additional features, akin to replicas of the dispersing mode  stemming from the same antiferromagnetic point and extend in energy up to $\sim$15-20 meV. These are both energy and momentum broadened beyond the spectrometer resolution and are not accounted for by linear spin wave theory.  We turn now to the interpretation of these high-energy spin dynamics shown in Figure \ref{fig:dispersion} $(c)$. As magnetic excitations above $\sim$ 2 meV are indicative of one dimensional correlations (Fig. \ref{fig:macs_lowT}), we model this dominant single magnon branch using an XXZ model (Fig. \ref{fig:dispersion} $(a)$) : \[ E(\vec{Q}) = 2JS\sqrt { (1+\delta_z)^2 - \gamma^2 (\vec{Q} )}\] where $ \gamma(\vec{Q}) = \cos(2\pi L)$ is the Fourier transform of the exchange coupling $J(\vec{Q})$ along the $c$-axis. $J$ (= 4.48 meV) and $(1+\delta_z)J$ are the nearest-neighbour interactions along the chain between $x$, $y$, and $z$ spin components. $\delta_z$~=~0.013 thus represents a weak exchange anisotropy along $z$, accounting for the energy minimum in Figure \ref{fig:dispersion} $(a)-(b)$. While one-dimensional fluctuations would be expected to destroy long-range three dimensional order with moment $g\langle S_{z}\rangle$, a finite $\delta_z$ protects the magnetic order from such devastating fluctuations. This is illustrated in the dependence of the spin reduction $\Delta S$, defined through $\langle S_{z}^{2} \rangle=(S-\Delta S)^{2}$, as a function of $\delta_z$ shown in the Supplemental Material for $S$~=~3/2. $\Delta S$ especially becomes large as $\delta_z$ goes to zero.

\begin{figure}
 \includegraphics[scale=0.34]{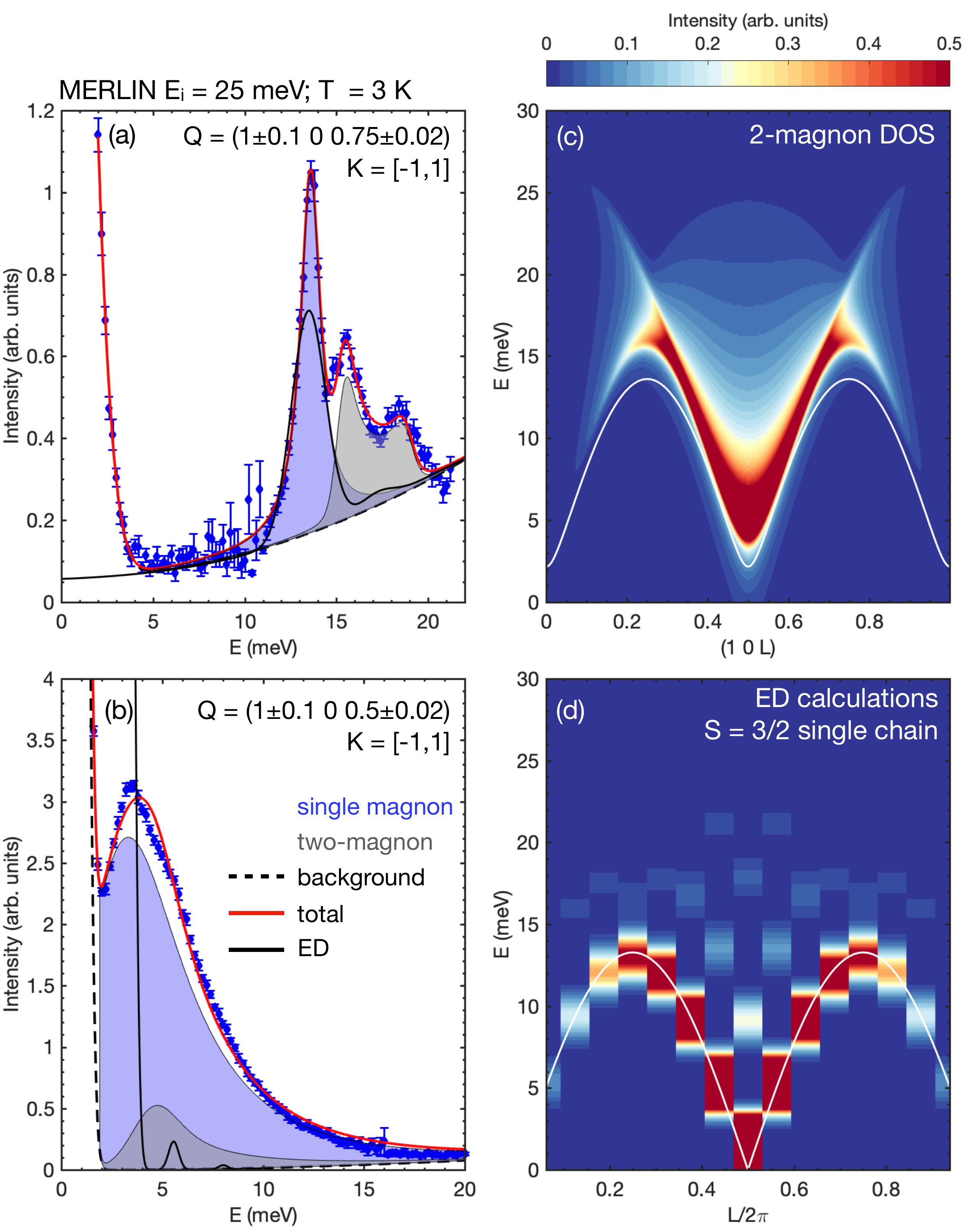}
 \caption{ \label{fig:2magnon} $(a)$-$(b)$ Constant-Q cuts around Q~=~(1~0~0.75) and Q~=~(1~0~0.5) respectively, from MERLIN. The shaded areas show the single magnon (blue) and the two-magnon contributions (grey); the black dashed line shows the background baseline and the red line is a fit comprising the total contribution of the one- and two magnon plus the background. The plain black line in $(a)$ corresponds to ED calculations. $(c)$ Calculated two-magnon density of states based on the XXZ dispersion measured in Fig. 1 $(a)$. $(d)$ Exact diagonalization calculation for a single S = 3/2 chain. The white curve is the single magnon mode as described in the text. }
\end{figure}

We now discuss the momentum and energy broadened features at 15-20 meV which qualitatively appear as replicas of the transverse single magnon spin waves discussed above, suggestive of an origin from multiparticle processes. To reproduce these features, we calculate the two-magnon density of states~\cite{Lovesey:book,Huberman2005} as detailed in the Supplemental Information. Such processes involve scattering from two magnons and are constrained by spin, momentum, and energy conservation. This causes an additional neutron cross-section, which is longitudinally polarized, in a wide region in energy and momentum, determined by the single magnon dispersion. The energy and momentum dependence of the kinematically allowed multimagnon scattering is shown in Figure \ref{fig:2magnon} $(c)$. The calculation shows a broad continuum extending up to twice the energy bandwidth of the single magnon mode (white curve), with two distinct features at the zone boundaries around 15 and 20 meV. A fit including the single magnon and the calculated two-magnon contributions at $\vec{Q}$=(1~0~0.5) and (1~0~0.75) is presented in Figure \ref{fig:2magnon} $(a)$-$(b)$ and is in agreement with the experiment.

To understand the distribution of intensity between the elastic, and inelastic one- and two-magnon channels, we compare the spectral weights with the zeroeth sum rule~\cite{Hohenberg1974} by calibrating the intensity from MERLIN (Fig. \ref{fig:dispersion} $(c)$) using Cr as an internal incoherent standard~\cite{Sears1992, Sarte2018,Xu2013}. The total moment sum rule of neutron scattering defines the integral of all spectral weight as : 
\[\int d^3\mathbf{Q} \int d\omega S(\mathbf{Q},\omega) = N S(S+1)=N \times 3.75 \] with $N$=2 the number of Cr$^{3+}$ ($S$~=~3/2) ions in a unit cell. The measured total integral including an elastic contribution of 0.6 (1) is 3.4(2). The elastic contribution was extracted from the integrated intensity of the first magnetic Bragg peak. The single crystal elastic value of $g \langle S_{z} \rangle=1.56$ is reduced in comparison to the full $gS$ value and the refined powder data~\cite{Damay2010} and is consistent with the presence of strong longitudinal fluctuations. Similar discrepancies between powder and single crystal data have been reported in low dimensional magnets~\cite{Stock2010}, but may also arise from energy integration in the powder diffraction experiment, thus taking into account spin fluctuations in the refined ordered magnetic moment.  The two-magnon component was estimated to be 0.9(2) by removing the elastic and single-magnon components. These measured integrals are compared against calculations for a $S$~=~3/2 chain in Table \ref{tab:sumrule}.  All spectral weight is accounted for and the energy and momentum broadened replica of the one-magnon mode is consistent with low-dimensional quantum fluctuations.

\begin{table}[t]
\caption{\label{tab:sumrule} Sum rules for the different components of the scattering evaluated for S = 3/2.}
\begin{ruledtabular}
\begin{tabular}{llll}
  &  Theory & Experiment \\
 \hline
$S(S+1)$        & 15/4=3.75  & 3.4(2) \\
$\Delta S$      & 0.53       & 0.5(1)\\
$\langle S_z^2\rangle$ & $(S-\Delta S)^2 = $ 0.94 & 0.6(1) = $(1.56/g)^2$ \\ 
Two-magnon      & $\Delta S (\Delta S+1)=$ 0.81    & 0.9(2) \\
One-magnon      & $(S-\Delta S)~(1 + 2\Delta S)=$2 & 1.9(1)  \\
\end{tabular}
\end{ruledtabular}
\end{table}

Finally, to compare our data to the exact model for a single S = 3/2 chain, exact diagonalization (ED) calculations were performed and the results are shown in Figure \ref{fig:2magnon} $(d)$. The model is based on an isotropic Heisenberg Hamiltonian with nearest-neighbor antiferromagnetic interactions and the exact calculations were performed for the longitudinal component $S^{zz}(\vec{Q},E)$ (details are provided in the Supplemental Information). The ED calculations have been normalized so that the integrated intensity obeys the partial moment sum rule and is equal to $S(S+1)/3$ (since the model is isotropic, the transverse components $S^{xx}$ and $S^{yy}$ carry the same spectral weight). As displayed in Figure \ref{fig:2magnon} $(d)$, this spectrum is characterized by a low energy boundary, which bears strong similarities with a single magnon mode. Its ``dispersion'' is fitted using the relation $E(L) = \alpha J_{ED} |\sin(2\pi L)|$, where $\alpha$~=~3.69 is a numerical constant, and $J_{ED}$~=~3.6~meV was determined so that $E(L)$ coincides with the experimentally observed magnon mode. Significant spectral weight is also found above this boundary, in a large region in energy and momentum. Strikingly, when comparing both models around the antiferromagnetic zone center, the longitudinal fluctuations calculated from our two-magnon model are found of the same order of magnitude as the longitudinal fluctuations expected in a quantum spin-3/2 chain from exact diagonalization. This hence further supports the idea that the low energy sector is governed by the magnetic order, while the high energy originates from quantum fluctuations. However, as shown on Figure \ref{fig:2magnon} $(a)$, where the ED calculation is compared against the two-magnon model and the experimental data, the isotropic model does not quite capture the high energy intensity at the zone boundary.

In the N\'eel state, $\beta$-CaCr$_{2}$O$_{4}$ displays spatially long-range magnetism with harmonic spin waves at low energy transfers, reminiscent of large-$S$ systems and parameterized here semiclassically.  This contrasts with the response at higher energies  which is momentum and energy broadened, resembling quantum fluctuations expected from low-dimensional $S$~=~1/2 magnets~\cite{Schulz1996, Essler1997, Lake2005, Lake2010, Zheludev2002, Zheludev2002bis, Zheludev2003} that disrupt spatially long-range order.  This is further confirmed here through analysis of the spectral weight, kinematics, and exact diagonalization calculations.  $\beta$-CaCr$_{2}$O$_{4}$ therefore displays a crossover from classical spin-waves to quantum fluctuations with the crossover energy defined by the local anisotropy.

It has  been shown that a non-collinear spin structure can enhance the two-magnon intensity through the coupling between transverse and longitudinal terms as theoretically studied in S~=~1/2 and S~=~3/2 non-collinear triangular magnets~\cite{Chernyshev2006,Chernyshev2009,Mourigal2013bis}. However, in the latter case, the predicted spectral weight related to multimagnon scattering is much weaker than the intensity observed in $\beta$-CaCr$_2$O$_4$, as also recently reported in another high spin non-collinear magnet \cite{Songvilay2018}. We therefore do not associate such mixing as the origin of the quantum fluctuations in $S$~=~3/2 $\beta$-CaCr$_{2}$O$_{4}$.

The static magnetism of $\beta$-CaCr$_{2}$O$_{4}$ displays multiple phases with a transition to a spin density wave state followed by cycloidal order at low temperatures~\cite{Damay2010}.  A thermally induced spin density wave is unusual given the insulating nature and hence a locally conserved moment is expected, unlike itinerant systems that display density wave phases and similar high energy dynamics~\cite{Rodriguez2011,Stock2014,Stock2015,Plumb2018}.  The close proximity of energetic longitudinal fluctuations originating from the chain nature may be the origin of this induced phase allowing these fluctuations to dominate critical fluctuations as suggested in the context of triclinic Cu$_{3}$Nb$_{2}$O$_{8}$~\cite{Nathan18:102}.  $\beta$-CaCr$_{2}$O$_{4}$ is therefore on the border between classical and quantum fluctuations dominating the phase transition and dynamics.

We suggest that the energy scale that protects the N\'eel order and the harmonic spin waves is determined by the local anisotropy set by the local crystalline electric field.  While octahedrally coordinated Cr$^{3+}$ is not expected to have any anisotropy given the lack of any orbital degree of freedom, local distortions of the crystalline electric field can mix in higher crystal field terms allowing anisotropic terms in the magnetic Hamiltonian~\cite{Yosida:book}. Given that classical N\'eel order in $\beta$-CaCr$_{2}$O$_{4}$ is stabilized by this anisotropy, it maybe that through pressure or disorder this phase can be suppressed giving way to a fully quantum disordered $S$~=~3/2 ground state. While, calculations for $S$~=~3/2 chains have found a strong sensitivity to disorder~\cite{Gil2002,Richter2020}, magnetic dilution has been investigated through chemical substitution in $\beta$-CaCr$_{2}$O$_{4}$ and has however shown that owing to inter-chain interactions, this system remains robust against disorder \cite{Songvilay2015}. 

We have experimentally and theoretically investigated the magnetic dynamics of $\beta$-CaCr$_{2}$O$_{4}$ to address the presence of single and two magnon processes.  The energy scale separating these dynamics is defined by a crystallographic anisotropy originating from the distorted environment around Cr$^{3+}$.  The dynamics in $S=3/2$ $\beta$-CaCr$_{2}$O$_{4}$ represents an example at the border between quantum and classical physics.

\begin{acknowledgments}
We acknowledge funding from the EPSRC, STFC and Investissements d'Avenir du LabEx PALM (ANR-10-LABX-0039-PALM). We acknowledge the support of the National Institute of Standards and Technology, U.S. Department of Commerce, and LLB-Orph\'ee (CEA Saclay, France) in providing the neutron research facilities used in this work. Access to MACS was provided by the Center for High Resolution Neutron Scattering, a partnership between the National Institute of Standards and Technology and the National Science Foundation under Agreement No. DMR-1508249. The work at Rutgers University was supported by the DOE under Grant No. DOE: DE-FG02-07ER46382.
\end{acknowledgments}

\bibliography{CaCrO}

\end{document}



\title{Supplemental information for ``From single- to two-magnon excitations in the S = 3/2 magnet $\beta$-CaCr$_2$O$_4$'' }

\author{M. Songvilay}
\affiliation{School of Physics and Astronomy, University of Edinburgh, Edinburgh EH9 3JZ, United Kingdom}
\altaffiliation{present address: Institut N\'eel, CNRS \& Univ. Grenoble Alpes, 38000 Grenoble, France}

\author{S. Petit}
\affiliation{Laboratoire L\'{e}on Brillouin, CEA-CNRS UMR 12, 91191 Gif-Sur-Yvette Cedex, France}

\author{F. Damay}
\affiliation{Laboratoire L\'{e}on Brillouin, CEA-CNRS UMR 12, 91191 Gif-Sur-Yvette Cedex, France}

\author{G. Roux}
\affiliation{Universit\'e Paris-Saclay, CNRS, LPTMS, 91405 Orsay, France}

\author{N. Qureshi}
\affiliation{Institut Laue-Langevin, 6 rue Jules Horowitz, Boite postale 156, 38042 Grenoble, France}

\author{H. C. Walker}
\affiliation{ISIS Neutron and Muon Source, Rutherford Appleton Laboratory, Chilton, Didcot OX11 0QX, United Kingdom}

\author{J. A. Rodriguez-Rivera}
\affiliation{NIST Center for Neutron Research, National Institute of Standards and Technology, 100 Bureau Drive, Gaithersburg, Maryland, 20899, USA}
\affiliation{Department of Materials Science, University of Maryland, College Park, Maryland 20742, USA}

\author{B. Gao}
\affiliation{Rutgers Center for Emergent Materials and Department of Physics and Astronomy, Rutgers University, 136 Frelinghuysen Road, Piscataway, New Jersey 08854, USA}

\author{S.-W. Cheong}
\affiliation{Rutgers Center for Emergent Materials and Department of Physics and Astronomy, Rutgers University, 136 Frelinghuysen Road, Piscataway, New Jersey 08854, USA}

\author{C. Stock}
\affiliation{School of Physics and Astronomy, University of Edinburgh, Edinburgh EH9 3JZ, United Kingdom}


\date{\today}

\pacs{}

\maketitle


\section{Structural characterisation and magnetic susceptibility}

\begin{figure} 
 \includegraphics[scale=0.33]{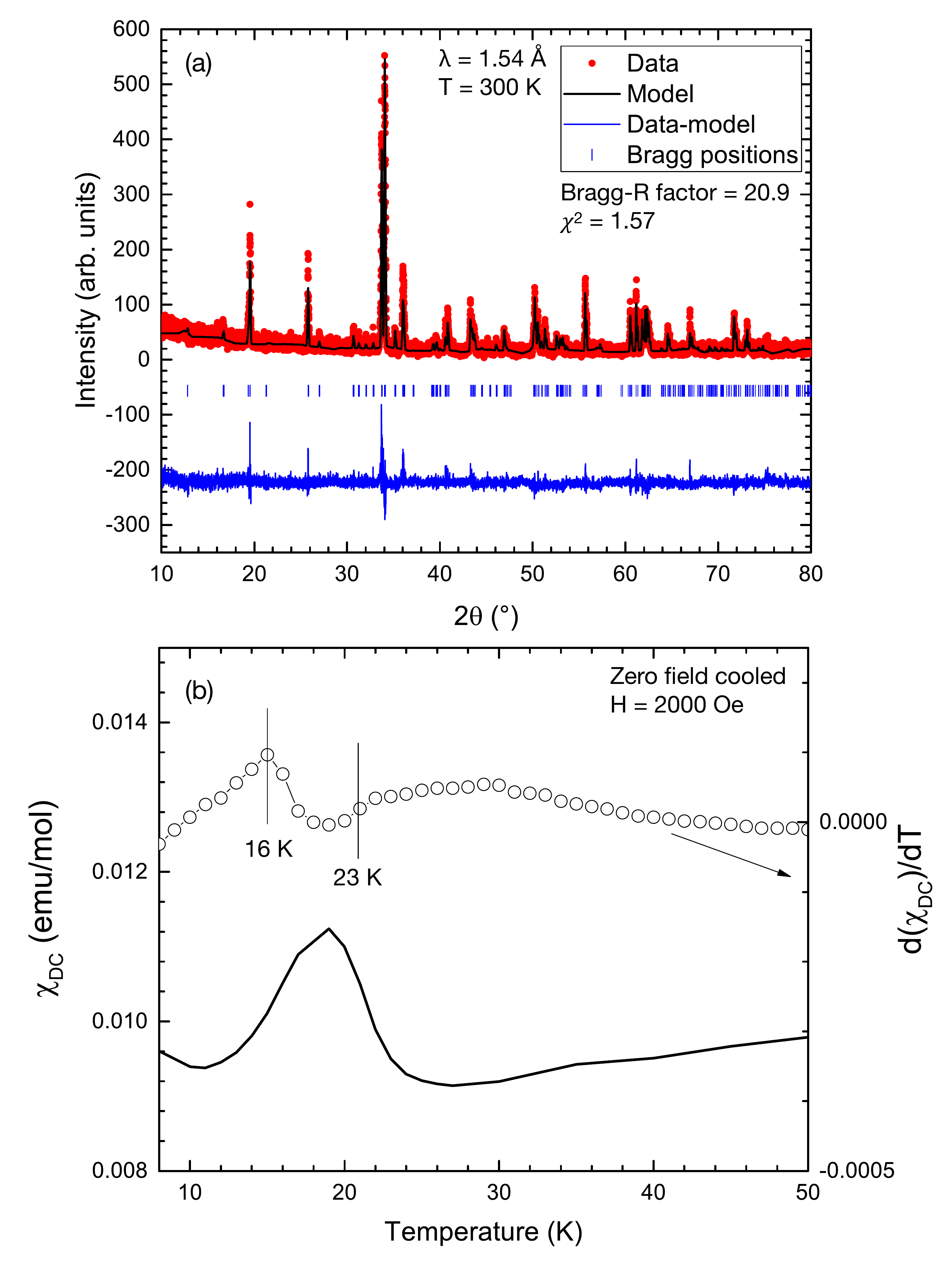}
 \caption{ \label{fig:diff} $(a)$ Structural Rietveld refinement of X-ray powder diffraction data measured on a crushed single crystal ($\lambda$ = 1.54 \AA). $(b)$ Magnetic DC susceptibility curve as a function of temperature.}
\end{figure}

X-ray powder diffraction was performed on a crushed crystal in order to check for any structural disorder. As shown in Figure \ref{fig:diff}, a Rietveld refinement was performed using the structural parameters of \cite{Damay2010}, which is in good agreement with the data. Moreover, there is no sign of broadening which would indicate the presence of structural disorder. This is further confirmed with susceptibility data measured on a single crystal, which displays two magnetic transitions, in agreement with the previously published susceptibility results \cite{Damay2010}.

\section{Neutron experiments}

\begin{figure}
 \includegraphics[scale=0.5]{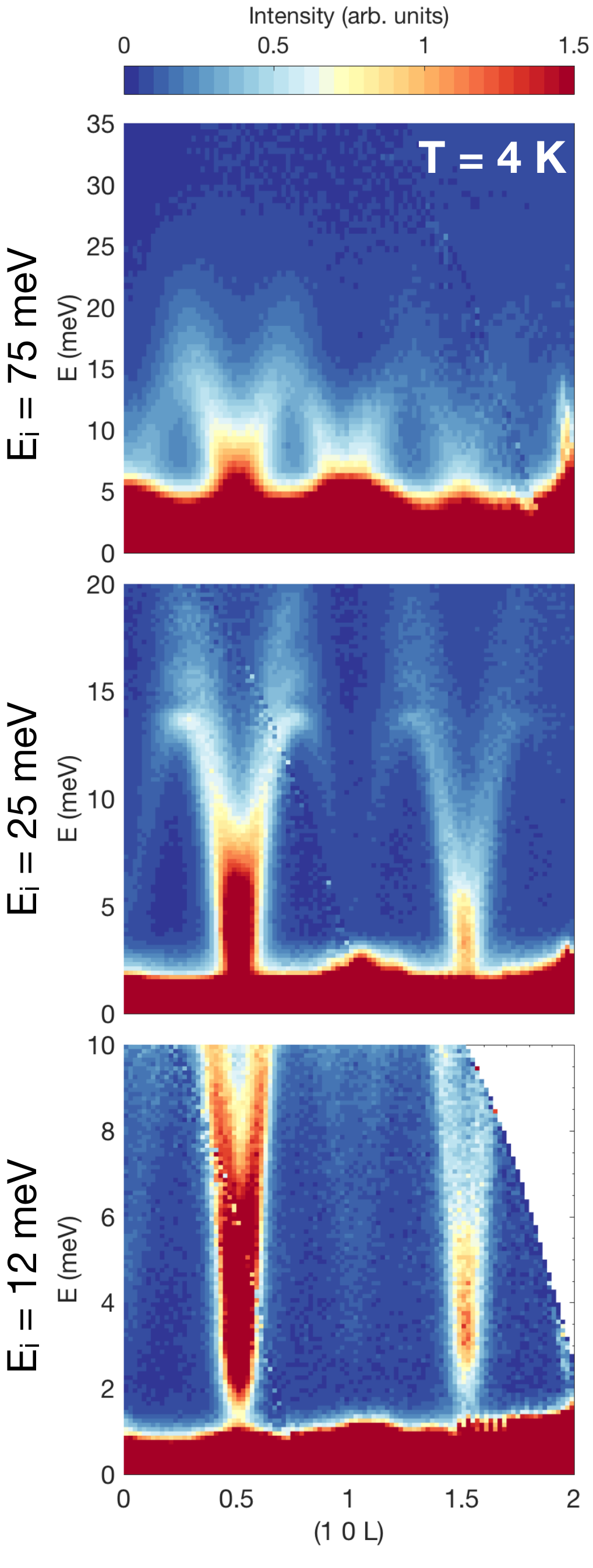}
 \caption{ \label{fig:merlin} Excitation spectrum of $\beta$-CaCr$_2$O$_4$ measured at T~=~4K on MERLIN using the multirep option allowing a simultaneous access to the incident energies E$_i$~=~75, 25 and 12~meV. The intensity as a function of momentum follows the magnetic form factor associated to Cr$^{3+}$ and therefore shows that the excitations are purely magnetic.}
\end{figure}

Neutron inelastic spectroscopy was performed on the thermal triple-axis spectrometer 2T1 (LLB-Orph\'ee, France), with constant E$_f$~=~14.7~meV using two PG filters between the sample and the analyser to remove higher order contamination. Low-energy excitations were investigated on the cold multi-detector triple-axis spectrometer MACS (NIST, US) with constant energies of E$_f$~=~5~meV using a Be filter between the monochromator and the sample and E$_f$~=~2.5~meV using a Be filter both between the monochromator and the sample and between the sample and the analyser. Time-of-flight neutron scattering measurements were performed on the MERLIN chopper spectrometer (ISIS, UK). By using a gadolinium fermi chopper spinning at 250 Hz in multirep mode with a corresponding disc chopper spinning at 50 Hz, incident energies of E$_i$ = 75, 25 and 12~meV were simultaneously used, as shown in Figure \ref{fig:merlin}. A $t0$ chopper was spun at 50 Hz to remove high energy neutrons.  The different experimental configurations with the calculated energy resolutions at the elastic energy (E=0 meV) are tabulated below.

\begin{table}[ht]
\caption{Spectrometer configurations and calculated energy resolutions.}
\begin{tabular} {c|c|c}
\hline
Instrument & E$_{final/initial}$ & $\delta$ E(0 meV) (FWHM) \\
\hline
 MACS (NIST) 	& 	E$_{final}$=2.5 meV 	& 0.08 meV  \\
 MACS  			& 	E$_{final}$=5.0 meV 	& 0.28 meV  \\
 \hline
 2T1 (LLB) 	& 	E$_{final}$=14.7 meV 	&  1.6 meV \\
 \hline
 MERLIN (ISIS)			& 	E$_{initial}$=12 meV 	&  0.24 meV \\
 MERLIN  			& 	E$_{initial}$=25 meV 	&  0.61 meV \\
 MERLIN  			& 	E$_{initial}$=75 meV 	&  2.9 meV \\
\hline
\end{tabular}
\label{table_pos}
\end{table}

\section{Application of CRYOPAD}

In the main text,  we reported the presence of two elastic peaks which disappeared at temperatures above T$_{N}$.  One was located at the incommensurate position and agreed with neutron powder diffraction while the other peak was located at $\vec{Q}$=(-1.5~0~$\sim$0.55). However, in the inelastic channel there were no gapless excitations emanating from this wavevector. We therefore applied neutron polarization analysis using {\sc cryopad} to check the magnetic nature of this peak on the D3 diffractometer (ILL, France).  The measured polarization matrix at $\vec{Q}$=(-1.5~0~$\sim$0.55) was found to be defined by the following:

\begin{equation}
P_{(-1.5~0~0.55)} = 
\begin{pmatrix}
{\bf{-0.23(18)}} & 0.5(3) &  -0.13(16) \\
-0.34(15) & 0.32(19) & 0.2(2) \\
0.2(2) & 0.07 (17) &  -1.3(3)
\end{pmatrix} \nonumber
\end{equation}

\noindent Of particular concern for this study is that the first matrix element $P_{11} \equiv P_{xx}$ (highlighted in the above matrix) must be equal to -1 for a purely magnetic peak given that this component represents the magnetic component perpendicular to the momentum transfer. This is the case for the magnetic Bragg peak $\vec{Q}$=(1~0~0.47) as $P_{xx}$=-0.99(2). The fact that $P_{xx} \neq -1$ for the $\vec{Q}$=(-1.5~0~$\sim$0.55) peak therefore indicates that the peak's origin is not purely magnetic and has a strong structural component.  Given the lack of any inelastic component originating from this peak we have not considered it in our analysis.

\section{Low-energy constant-E cuts}

Constant-energy cuts through the magnetic dispersion were performed at low energy using the MACS spectrometer with constant E$_f$~=~5 meV. The position in momentum of the magnetic mode was extracted by fitting the data to Gaussian functions and the extracted positions are given in Table \ref{tab:fitL}.

\begin{table}
\caption{Fitted position along (1~0~L) of the magnetic excitation at several energy transfers.}
\begin{tabular} {c|c|c}
\hline
Energy  & Position peak 1 (r.l.u) & Position peak 2 (r.l.u) \\
\hline
1 meV      & 0.478(1) & - \\
1.5 meV   & 0.479(2) &  0.51(3) \\
1.75 meV & 0.485(2) & 0.52(3) \\
2.2 meV   & -             & 0.501(5) \\
\hline
\end{tabular}
\label{tab:fitL}
\end{table}

\section{Dimensionality of low-energy spin fluctuations}

\begin{figure} [b!]
 \includegraphics[scale=0.3]{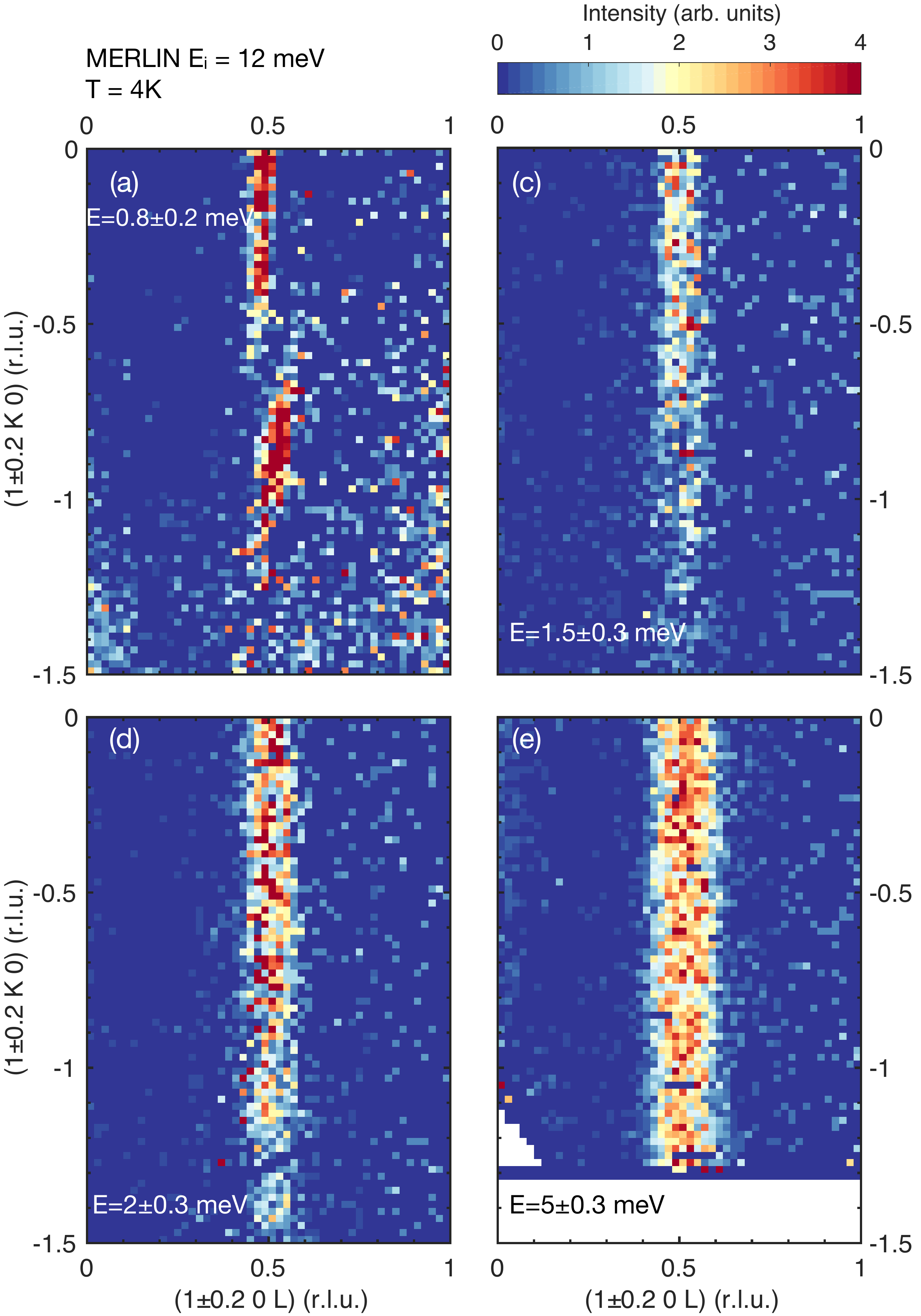}
 \caption{ \label{fig:dimen} Constant energy slices from MERLIN (E$_{i}$=12 meV) in the (1 K L) scattering plane.  The data show little correlations along the crystallographic $b$-axis.}
\end{figure}

In the main paper we discussed the crossover from two dimensional to one dimensional spin correlations in $\beta$-CaCr$_{2}$O$_{4}$.  The low-energy magnetic fluctuations were described by a linear spin-wave theory given the underlying cycloidal ground state.  At higher energies, above energies defined by the anisotropy, the spin fluctuations were found to originate from quantum chain-like fluctuations.  The data in the main paper were restricted to scattering in the (H 0 L) scattering plane due to expectations that the coupling along the $b$-axis is weak, consistent with isostructural CaFe$_{2}$O$_{4}$. This is confirmed in Fig. \ref{fig:dimen} that plots a series of constant energy slices in the (1 K L) scattering plane.  At energies above at least 1 meV (Fig. \ref{fig:dimen} $(b)-(d)$), the scattering forms rods along K indicative of a lack of dynamic correlations.  We therefore conclude that there is little coupling along $b$ and that the low-energy spin-waves are two dimensional in nature as described in the main text.

\begin{figure} [b!]
 \includegraphics[scale=0.4]{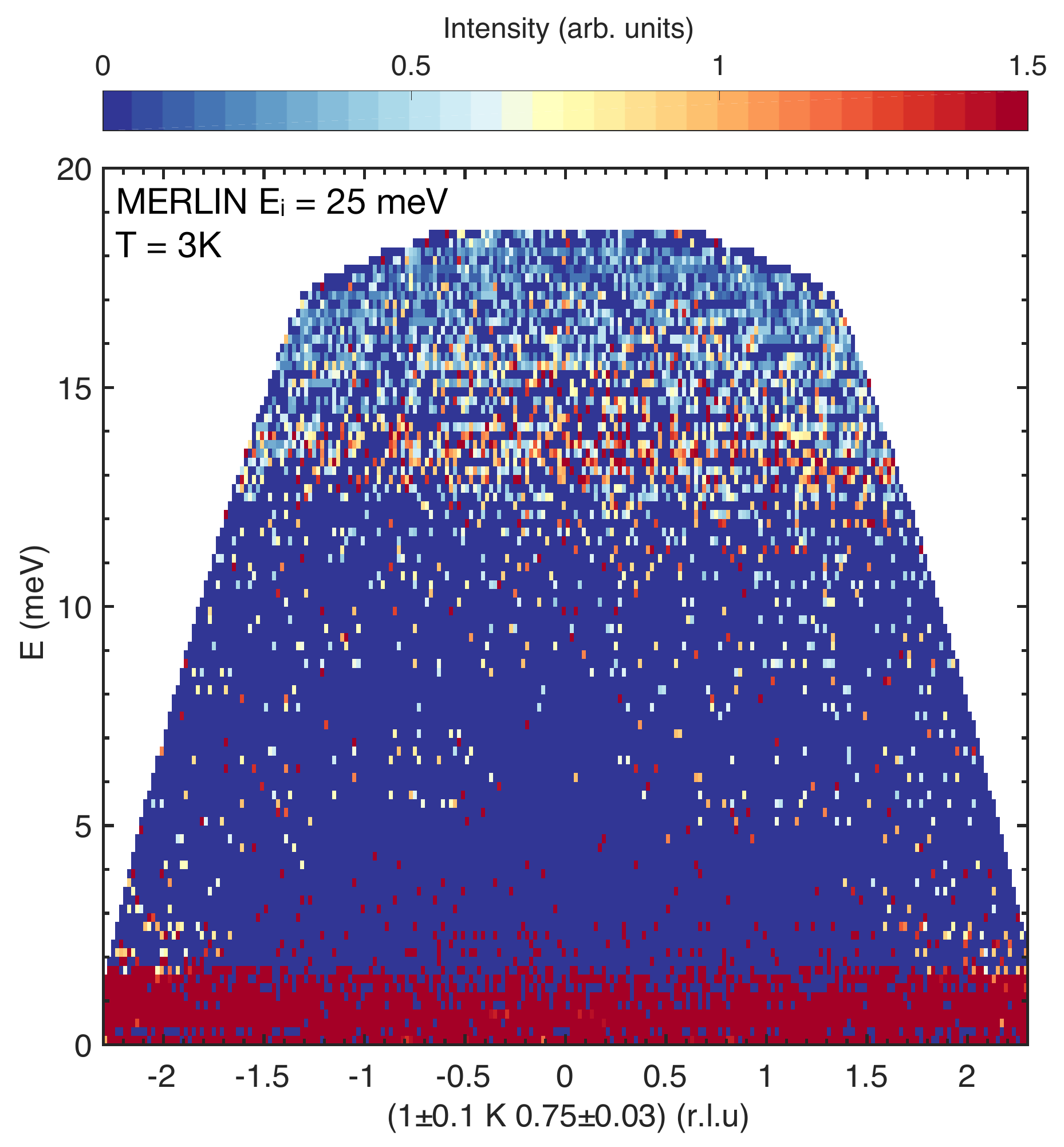}
 \caption{ \label{fig:1K0p75} Constant-Q slice from MERLIN (E$_{i}$=25 meV) showing the momentum dependence of the excitation spectrum in the (1~K~0.75) direction at T = 3 K.}
\end{figure}

Figure \ref{fig:1K0p75} shows the excitation spectrum in the (1~K~0.75) direction. The intensity associated to the two-magnon scattering observed above 12 meV is momentum independent in this direction, and therefore justifies the large range in K = \lbrack -1,1 \rbrack chosen for the data integration in Figures 1 and 4 of the main article. Indeed, this shows that the two features observed in the continuum above 12 meV does not result from the integration in the K direction.

\section{Spin wave calculations \label{appendix-sw}}

\begin{figure}
 \includegraphics[scale=0.2]{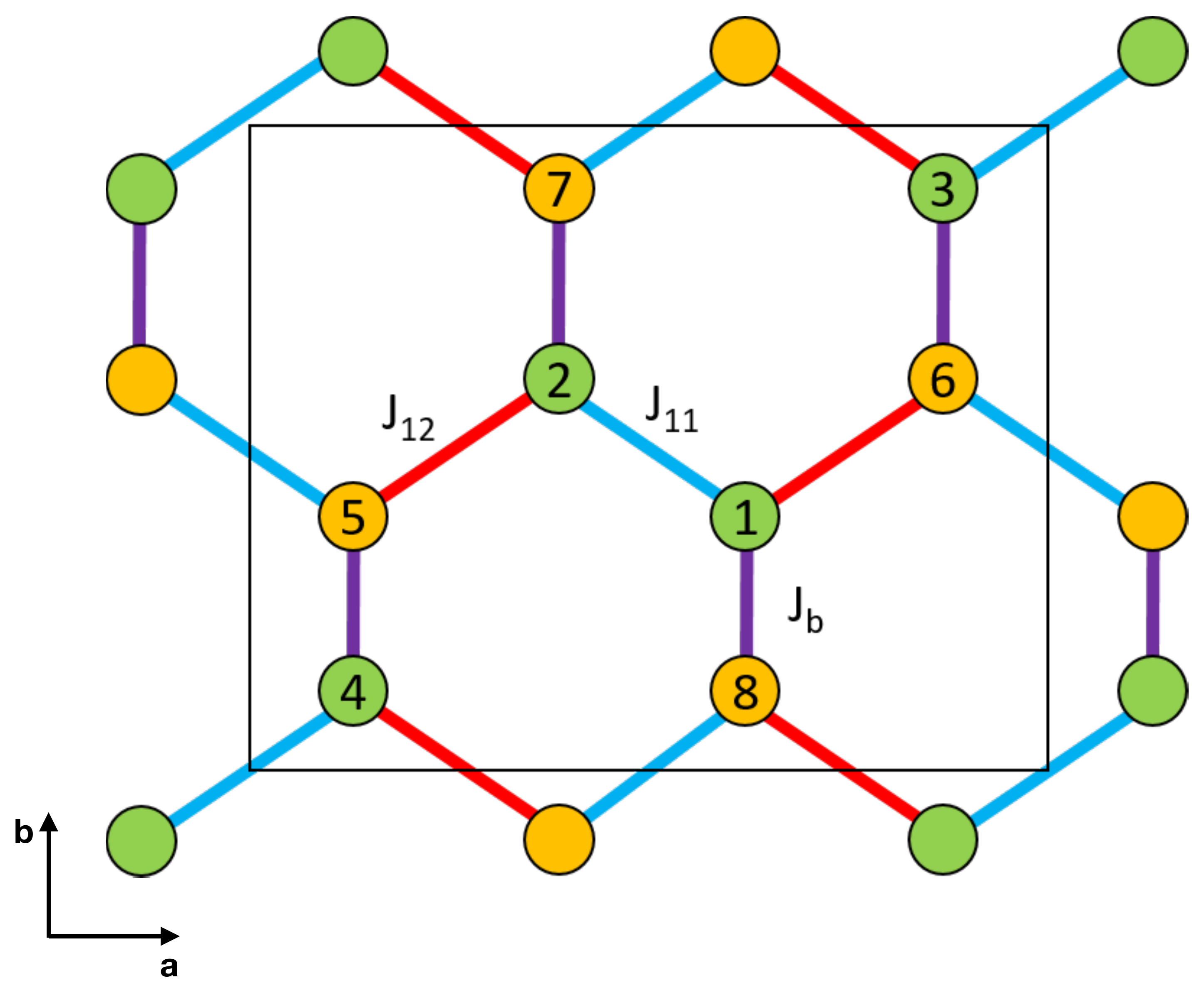}
 \caption{ \label{sw1} Sketch of the unit cell in the ($a,b$) plane. Green and yellow circles denote the Cr$_1$ and Cr$_2$ spins. Exchange interactions are indicated by links of different colors, highlighting the honeycomb-like structure of the connectivity.}
\end{figure}

The Cr$^{3+}$ spins occupy two distinct crystallographic positions Cr$_1$ and Cr$_2$, both on Wyckoff position 4c $(x, y, 1/4)$ of the $Pbnm$ space group. Those spins form chains running along the c-axis. The crystal symmetry is such that for any given chain in the lattice, the Cr positions in any adjacent chain are shifted by $c/2$. The magnetic lattice is therefore made out of chains interconnected to create a honeycomb-like network in the ($a,b$) plane. As a result, each spin has two intra-chain nearest neighbors and 6 inter-chain nearest neighbors. According to this structure, one should distinguish the intra-chain $J$, and the inter-chain couplings $J_{11}$, $J_{22}$, $J_{12}$ and $J_b$, as illustrated in Figure \ref{sw1} (for the sake of simplicity, $J_{11}=J_{22}$ is assumed in the following).

\begin{figure}
\includegraphics[scale=0.43]{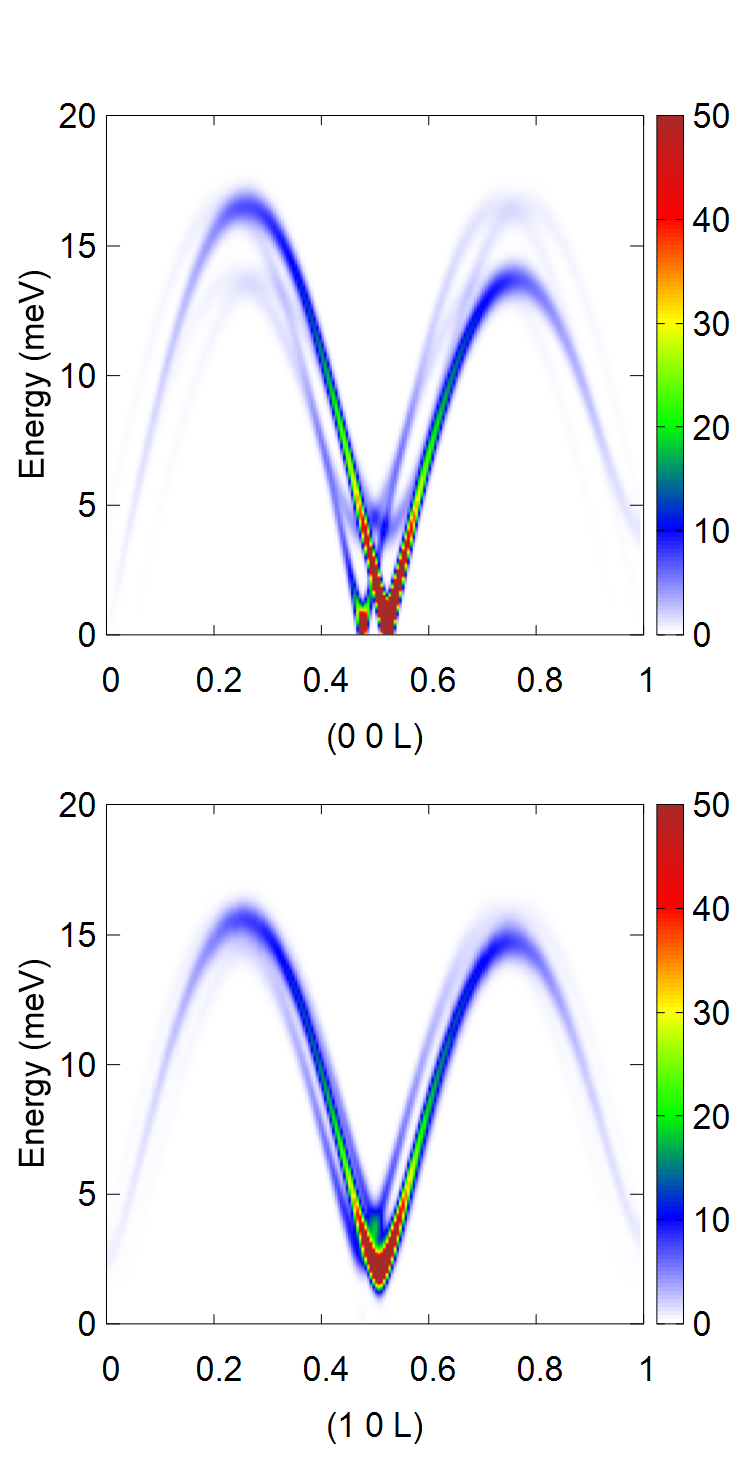}
\caption{\label{sw2} Spin wave spectrum calculated in the case of a uniform chiral pattern. The two acoustic in plane modes are observed along (0 0 L), while the out of plane mode is observed along (1 0 L).}
\end{figure} 

Even if it {\it  does not} correspond to the actual structure, it is instructive to consider the case where all the spins rotate the same way. Straightforward calculation shows that the energy per spin is then given by:
\begin{equation}
E = 2J \cos 2 \pi k + 2(J_{11}+J_b+J_{12}) \cos 2 \pi k/2
\label{E1}
\end{equation}
where $k$ is the propagation vector. Minimization with respect to $k$ yields:
\begin{equation}
(J_{11}+J_b+J_{12}) = - 4J \cos \pi k
\end{equation}
Writing $k=1/2 +\eta$, with $\eta << 1$ ($\eta$ is the incommensurability parameter), we obtain:
\begin{equation}
\eta \approx \frac{1}{2\pi} \frac{J_{11}+J_b+J_{12}}{2J}
\end{equation}
Spin dynamics calculations performed with the {\sc spinwave} software, developed at LLB \cite{Petit-SW}, yield the spectrum displayed in Figure \ref{sw2}. Using $k=0.477$, we find $J=$ 4.5 meV and $J_{11}=J_{12}= J_b \approx$ 0.48 meV. As it is clear from the comparison with experiments, this simple spin wave model fails to capture the high energy part of the INS data observed in CaCr$_2$O$_4$.

\begin{figure}
 \includegraphics[scale=0.2]{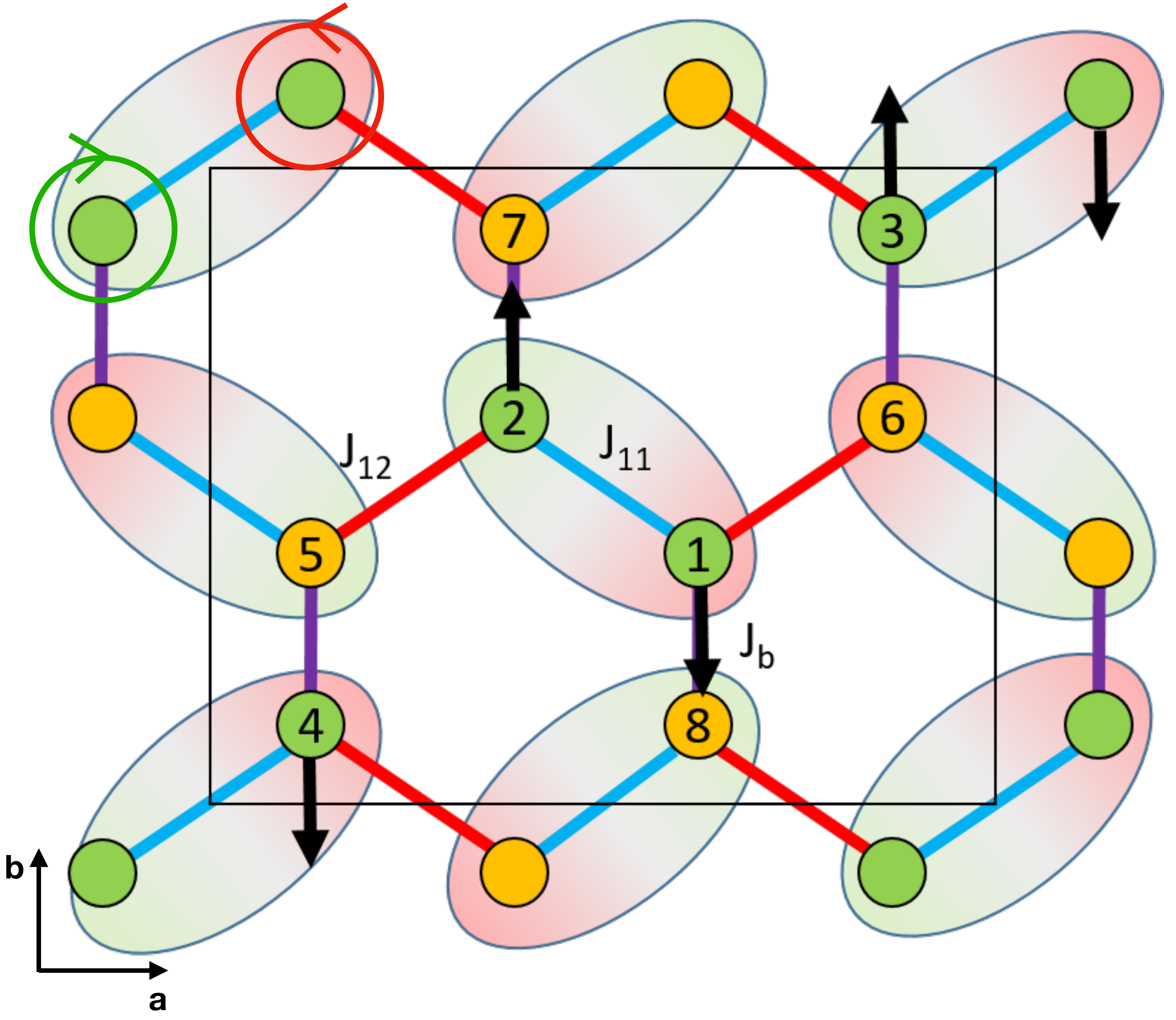}
 \caption{ \label{sw3} Sketch of the unit cell in the ($a,b$)  plane. The clockwise and anti-clockwise senses of rotations are indicated by green and red shaded areas. Black arrows show the directions of the DM vector D.}
\end{figure} 

A striking feature of the actual magnetic structure lies in its staggered chirality. As shown in Figure \ref{sw3}, the spins labeled 1, 6, 4 and 7 rotate clockwise, while spins 2, 5, 8 and 3 rotate anti-clockwise. This pattern form pairs of chains (coupled by $J_{12}$) rotating the same way, namely 1-6, 4-7, 2-5 and 8-3. As proposed in \cite{Damay2010}, the Dzyaloshinskii-Moriya (DM) interaction may be the driving force able to stabilize this particular chiral pattern. These authors argue that there is a direct competition between the weak $J_{11}, J_{12}, J_b$ interactions, that favor uniform chirality in a given pair of adjacent chains, and a DM interaction that favors opposite chiralities. The DM term writes $D.(S_i \times S_j)$ where the indexes $i$ and $j$ refer to first-neighbor ions along a chain. Furthermore, the $D$ vector is directed in the xy-plane and its $y$ component can couple to non-collinear spin arrangements in the ($x, z$) -plane. The crucial point is that the direction of the $D$ vector is reversed by the two-fold rotation (screw axis) along the z-axis that relates on one hand each pair of adjacent Cr$_1$-Cr$_1$ chains, and each pair of adjacent Cr$_2$-Cr$_2$ chains on the other hand. In other words, chains coupled by $J_{11}$ and $J_{22}$ have opposite D vectors. This can favor opposite sense of rotations in such pairs (see Figure \ref{sw3}). We thus write the spin coordinates at the different sites:
\begin{eqnarray*}
S_1(z) & = & (\cos 2\pi k z ,0, \sin 2\pi k z) \\
S_2(z) & = & (\cos 2\pi k(z+\frac{1}{2}) ,0, -\sin 2\pi k(z+\frac{1}{2})) \\
S_3(z) & = & (\cos 2\pi k z ,0, -\sin 2\pi k z) \\
S_4(z) & = & (\cos 2\pi k(z+\frac{1}{2}) ,0,  \sin 2\pi k(z+\frac{1}{2})) \\
S_5(z) & = & (\cos 2\pi k z ,0, -\sin 2\pi k z) \\
S_6(z) & = & (\cos 2\pi k(z+\frac{1}{2}) ,0,  \sin 2\pi k(z+\frac{1}{2})) \\
S_7(z) & = & (\cos 2\pi k z ,0, \sin 2\pi k z) \\
S_8(z) & = & (\cos 2\pi k(z+\frac{1}{2}) ,0, -\sin 2\pi k(z+\frac{1}{2}))
\end{eqnarray*}
hence the total energy per spin:
\begin{equation}
E = 2J \cos 2\pi k + D \sin 2\pi k + 2 J_ {12} \cos 2\pi k/2
\label{E2}
\end{equation}
Note that contributions from $J_{11}$ and $J_b$ vanish when summing over $z$. Minimization with respect to $k$ yields:
\begin{equation*}
D \cos 2\pi  k =  2J \sin 2\pi k + J_{12} \sin 2\pi k/2 
\end{equation*}
Using $k=1/2 +\eta$ and $\eta << 1$, we obtain:
\begin{equation}
\eta \approx \frac{1}{2\pi} \frac{D+J_{12}}{2J}
\end{equation}
Comparing Eq. \ref{E1} and Eq. \ref{E2}, we conclude that the staggered pattern is more favorable from an energetic point of view if $D \sin 2\pi k <  2 (J_{11}+J_b) \cos 2\pi k/2$, hence:
\begin{equation}
D >  J_ {11}+J_b
\nonumber
\end{equation}

\section{Multimagnon and longitudinal component of the scattering}

The momentum dependence of the magnetic scattering in CaCr$_{2}$O$_{4}$ at energy transfers above $\sim$ 5 meV is indicative of chain-like correlations. The energy and momentum broadened response measured both on MACS and MERLIN is analogous to multiparticle response reported previously in low dimensional magnets. In this section we consider the scattering from a chain and outline the momentum response and the amount of spectral weight in the multiparticle neutron cross section. \\ 

We consider the Hamiltonian for a spin chain with an anisotropy along the $z$ direction due to the distorted crystalline electric field resulting from the distorted oxygen octahedra surrounding each Cr$^{3+}$ ion. By applying perturbation theory, such a distortion can be written as an additional term in the Hamiltonian. The latter writes as an XXZ Hamiltonian:

\begin{equation}
H= J \sum_{i,j} \left[  \vec{S}_{i}\cdot \vec{S}_{j} + \delta_{z} S_{i}^{z}S_{j}^{z} \right].
\label{eq:alpha}
\nonumber
\end{equation}

\noindent The Hamiltonian can be converted to ladder operators $a$ and $a^{\dagger}$ through the Holstein Primakoff transformation:
\begin{eqnarray*}
S_x &\approx& \frac{\sqrt{S}}{2} (a+a^{\dagger}) \\
S_y &\approx& \frac{\sqrt{S}}{2i} (a-a^{\dagger}) \\
S_z &=& S- (a^{\dagger}a)
\nonumber
\end{eqnarray*}
and then diagonalized through the Bogoliubov transformation
\begin{equation}
a= u b + v b^{\dagger} 
\label{eq:alpha}
\nonumber
\end{equation}
with $u=\cosh \theta$, $v=\sinh \theta$, $\tanh 2\theta = -\gamma(\vec{Q})/(1+\delta_{z})$ and $\gamma(\vec{Q})=\cos (2\pi L)$. At this level of approximation, the ground state is the vaccum of the Bogoliubov particles, while the excitations are free bosons determined by the energy dispersion:
\begin{equation}
E(\vec{Q})=2JS \sqrt{\left[ (1+\delta_{z})^{2} -\gamma^{2}(\vec{Q}) \right]}
\label{eq:alpha}
\nonumber
\end{equation}

\begin{figure}
 \includegraphics[scale=0.38]{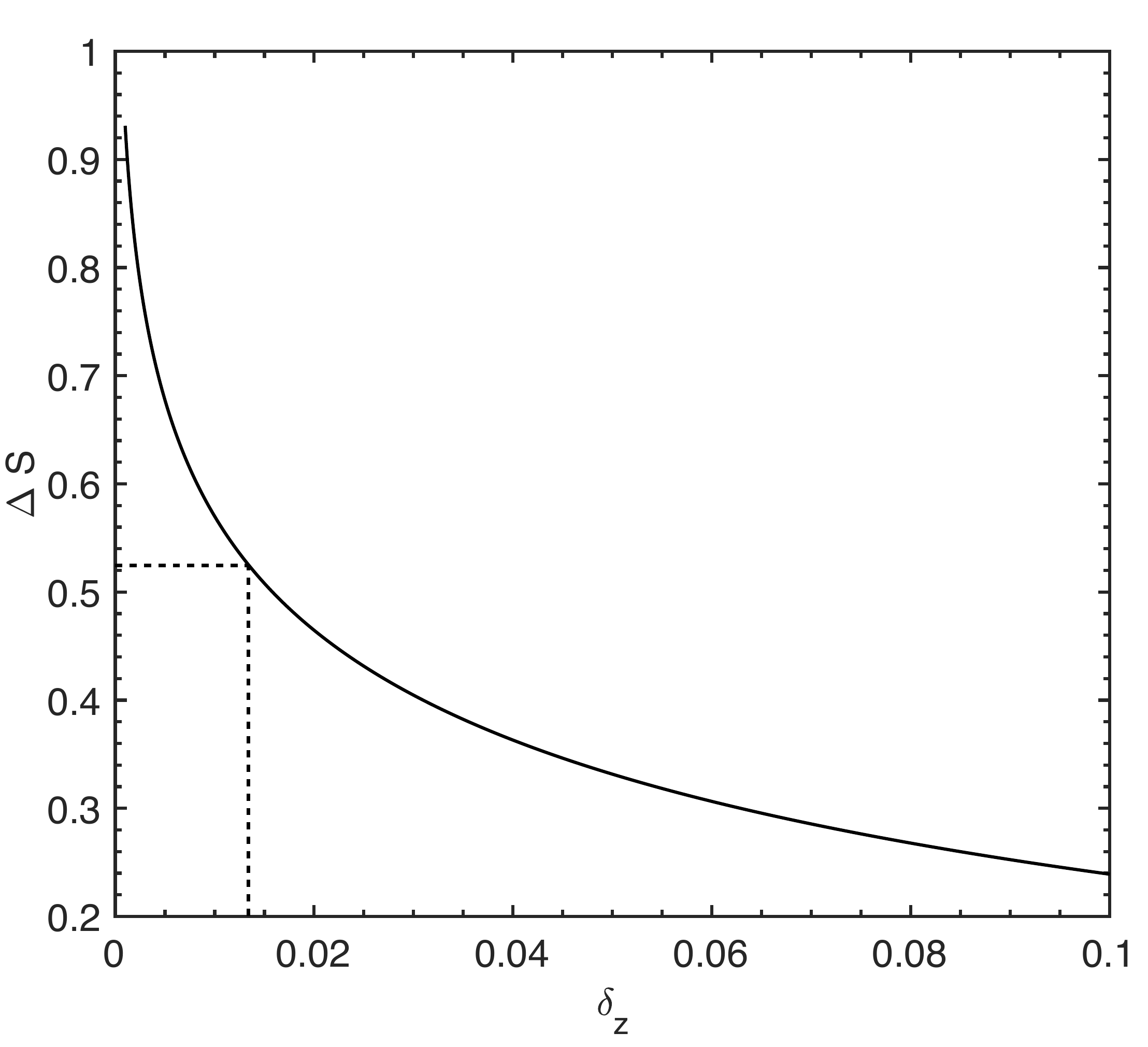}
 \caption{ \label{fig:deltaS} Evolution of the spectral weight $\Delta S$ for the longitudinal spin component as a function of the anisotropy $\delta_z$.}
\end{figure}

The amount of fluctuations is characterized by the expectation value of the $S_z$ component and determined from the Holstein Primakoff transformation,
\begin{equation}
\langle S_{z} \rangle=S-\langle a^{\dagger} a\rangle = S- \Delta S
\label{eq:alpha}
\nonumber
\end{equation}
\noindent
Applying the Bogoliubov transformation above, we can write the extra $\Delta S$ component as,
\begin{equation}
\Delta S = \frac{1}{N} \sum_{q}v_{q}^{2}.
\label{eq:alpha}
\nonumber
\end{equation}
\noindent This is calculated to be 0.53 for $\delta_{z}=0.013$ using the above dispersion to model the sharp transverse spin excitations in CaCr$_{2}$O$_{4}$ and performing the sum numerically. $\Delta S$ is sensitive to the value of $\delta_{z}$, as illustrated in figure \ref{fig:deltaS}.

Neutron scattering measures the dynamical correlation functions given by
\begin{equation}
{\cal S}^{ab}(\vec{Q},E) = \frac{1}{N} \sum_{i,j} e^{i\vec{Q}.(\vec{r}_i-\vec{r}_j)}\int dt~e^{i E t}~\langle S^a_i S^b_j(t) \rangle
\nonumber
\end{equation}
where N is the total number of spins and the sum runs over all sites i and j in the lattice. The transverse spin correlations (one magnon scattering) are given by:
\begin{equation}
{\cal S}^{xx}(\vec{Q},E)={\cal S}^{yy}(\vec{Q},E) = (S-\Delta S)~\frac{u_q^2+v_q^2}{2}~\delta(\omega-E(\vec{Q}))
\nonumber
\end{equation}
Integrating over energy and wavevector $\vec{Q}$, the total spectral weight contained in those transverse correlations writes: 
\begin{equation}
2(S-\Delta S)~\frac{1}{N}\sum_{q}\frac{u_q^2+v_q^2}{2} = (S-\Delta S)~\left(1 + 2\Delta S \right)
\nonumber
\end{equation}
The longitudinal spin correlations ${\cal S}^{zz}(\vec{Q},E)$ are given by: 
\begin{eqnarray*}
{\cal S}^{zz}(\vec{Q},E) &=& 
\int dt~e^{iE t} \sum_{i,j} e^{i\vec{Q}.(\vec{r}_i-\vec{r}_j)}~\\
& & \times \langle (S- a^{\dagger}_i a_i) \times (S- a^{\dagger}_j a_j)(t)\rangle
\nonumber
\end{eqnarray*}
On top of an elastic contribution (Bragg peak), ${\cal S}^{zz}$ especially encompasses a non-trivial contribution derived from a Hartree-Fock type expansion assuming no interactions and shown to be,
\begin{eqnarray*}
{\cal S}^{zz}(\vec{Q},E) &=&\frac{1}{N} \sum_{\vec{Q}_{1} \vec{Q}_{2}} f(\vec{Q}_{1},\vec{Q}_{2})~\\
& & \times \delta(\vec{Q}-\vec{Q}_{1}+\vec{Q}_{2})~ \delta(E-E_{\vec{Q}_{1}}-E_{\vec{Q}_{2}})
\label{eq:alpha}
\nonumber
\end{eqnarray*}
\noindent where $ f(\vec{Q}_{1},\vec{Q}_{2})={1\over 2} [u(\vec{Q}_{1})v(\vec{Q}_{2})-u(\vec{Q}_{2})v(\vec{Q}_{1})]^{2}$ as applied previously to Rb$_{2}$MnF$_{4}$ \cite{Huberman2005}. Integrating over energy and wavevector $\vec{Q}$, the total spectral weight contained in this contribution writes: 
\begin{equation}
\frac{1}{N}\sum_{q_1,q_2}\frac{1}{2} [u_1 v_2-u_2 v_1]^{2} = \Delta S \left(1 + \Delta S \right)
\nonumber
\end{equation}

Since the total spectral weight arising from the different contributions is $S(S+1)$, the spectral weight contained in the Bragg peak is given by: 
\begin{equation}
S(S+1) - (S-\Delta S) \left(1 + 2 \Delta S \right) - \Delta S \left(1 + \Delta S \right)  = (S-\Delta S)^2
\nonumber
\end{equation}

This same expression can be obtained through the application of the Hartree-Fock approximation to the formula {\cal S}$^{zz}$ above.  These different expressions are used in the Table 1 of the main text.

\section{Sum rules}

In order to study the two-magnon scattering contribution to the total spectral weight, the total moment sum rule was applied, by integrating the normalised inelastic intensity from the MERLIN data. The intensity, corrected from a constant background, was first converted into absolute units, using the sample incoherent elastic scattering, following the procedure described in \cite{Xu2013,Sarte2018}. For this, a constant-Q scan was extracted from the MERLIN data in a Q range containing no Bragg peak (around Q~=~(1~0~0.2)). The energy integrated intensity expresses as: $\int I_{incoh}(Q,E)dE~=~N~\sum_j~\sigma^{incoh}_j$ where $\sigma^{incoh}$ is the incoherent neutron scattering cross-section in barns of each atom. In our case, the incoherent cross-section of Ca and O being negligible, only Cr atoms are taken into account, with $N$=2 chromium atoms per unit cell. The calibration constant to normalise the intensity was therefore calculated as the following : $A=N\sigma^{incoh}(Cr)/\int I_{incoh}(Q,E)dE$. Using $\sigma^{incoh}(Cr)$ = 1.83 barns \cite{Sears1992}, N = 2, and $\int I_{incoh}(Q,E)dE$ = 13.25 counts.meV, the calibration constant is $A$~=~0.276~barns/count.meV.

The normalised intensity $A I(\textbf{Q},\omega) $ is proportional to the dynamic spin correlation function :

\[{\cal S}(\textbf{Q},\omega) = \frac{AI(\textbf{Q},\omega)}{\left( \frac{\gamma r_0}{2} \right)^2 (g\mu_B)^2 |f(\textbf{Q})|^2}    \]

where $ \left( \frac{\gamma r_0}{2} \right)^2$ equals 73 mb sr$^{-1}$, $g$ is the Land\'e factor equal to 2 for Cr$^{2+}$ and $f(\textbf{Q})$ is the magnetic structure factor calculated for Cr$^{2+}$. 
The dynamic structure factor then obeys the total moment sum rule:

\begin{equation}
\label{eq:sumrule}
\int d^3\mathbf{Q} \int d\omega {\cal S}(\mathbf{Q},\omega) = N S(S+1) 
\end{equation}

with $N$=2 the number of chromium atoms in a unit cell.

\begin{figure}
 \includegraphics[scale=0.32]{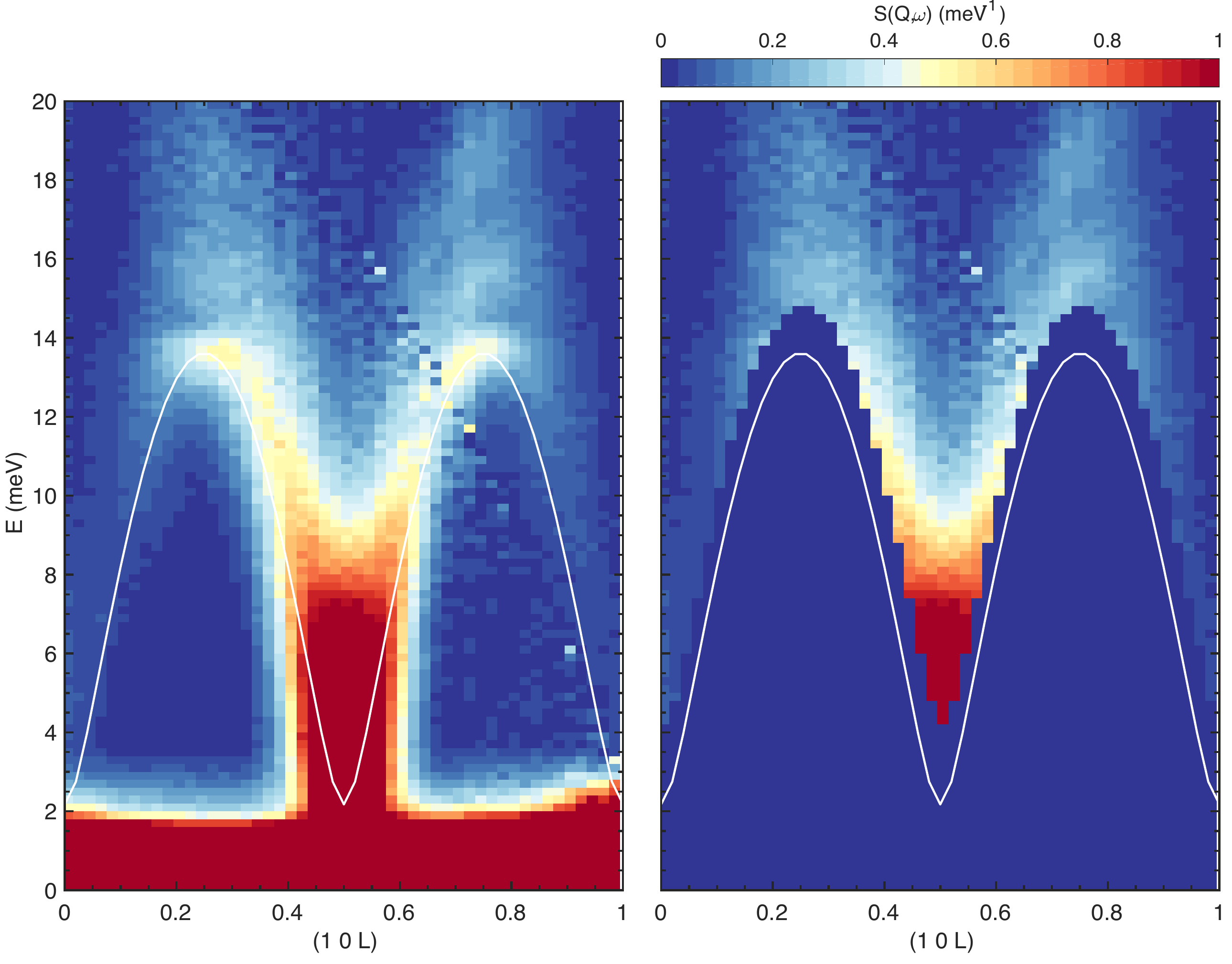}
 \caption{ \label{fig:sumrule} Two-magnon continuum isolated from the total spectrum by removing the single-magnon part.}
\end{figure}

As shown in figure \ref{fig:sumrule}, the continuum was isolated by removing the low-energy part of the spectrum, including the single-magnon mode. To perform this, the intensity below the single-magnon curve (shown as the white plain curve) was removed. The energy linewidth of the single-magnon mode was also taken into account by fitting the energy linewidth as a function of momentum from constant-Q scans, using an underdamped harmonic oscillator model:
\begin{equation}
{\cal S}(\vec{Q},\omega) = R(\omega)~I_0 \left(\frac{\Gamma_0}{\Gamma_0^2 + (\omega-\omega_0)^2} - \frac{\Gamma_0}{\Gamma_0^2 + (\omega + \omega_0)^2}\right)
\nonumber
\end{equation}
where $R(\omega)=(1 + n(\omega))$ is the detailed balance factor, $I_0$ is a constant, $\omega_0$ is the energy position of the single-magnon mode and $\Gamma_0$ is the single-magnon energy linewidth.

After removing the lower part of the spectrum, the remaining intensity was then integrated following equation (\ref{eq:sumrule}), which yields the value 0.9(2), close to the expected value of 0.8 extracted from the two-magnon calculation.

\section{Exact diagonalization results on the spin-$3/2$ Heisenberg chain}

We consider an antiferromagnetic spin chain model with either spin-$1/2$ or spin-$3/2$ atoms with positive nearest neighbour coupling $J$:

\begin{equation}
\Ham =  J\sum_{i=1}^L\mathbf{S}_i\cdot\mathbf{S}_{i+1} 
\end{equation}

where $\mathbf{S}_i$ is the spin operator at site $i$ and we write the total spin along the $z$-direction as $S^{z}_{\text{tot}}=\sum_iS_i^{z}$.

We use exact diagonalization calculations using both full diagonalization techniques and Lanczos calculations for calculations of thermodynamics and spectral functions. We fix $S^{z}_{\text{tot}}$ and use translational and spin inversion symmetries, which allows us, in particular, to obtain dispersion relation for the excitation. 

\subsection*{Dynamical structure factor and dispersion relations}

\begin{figure}[h]
\centering
\includegraphics[width=0.9\columnwidth,clip]{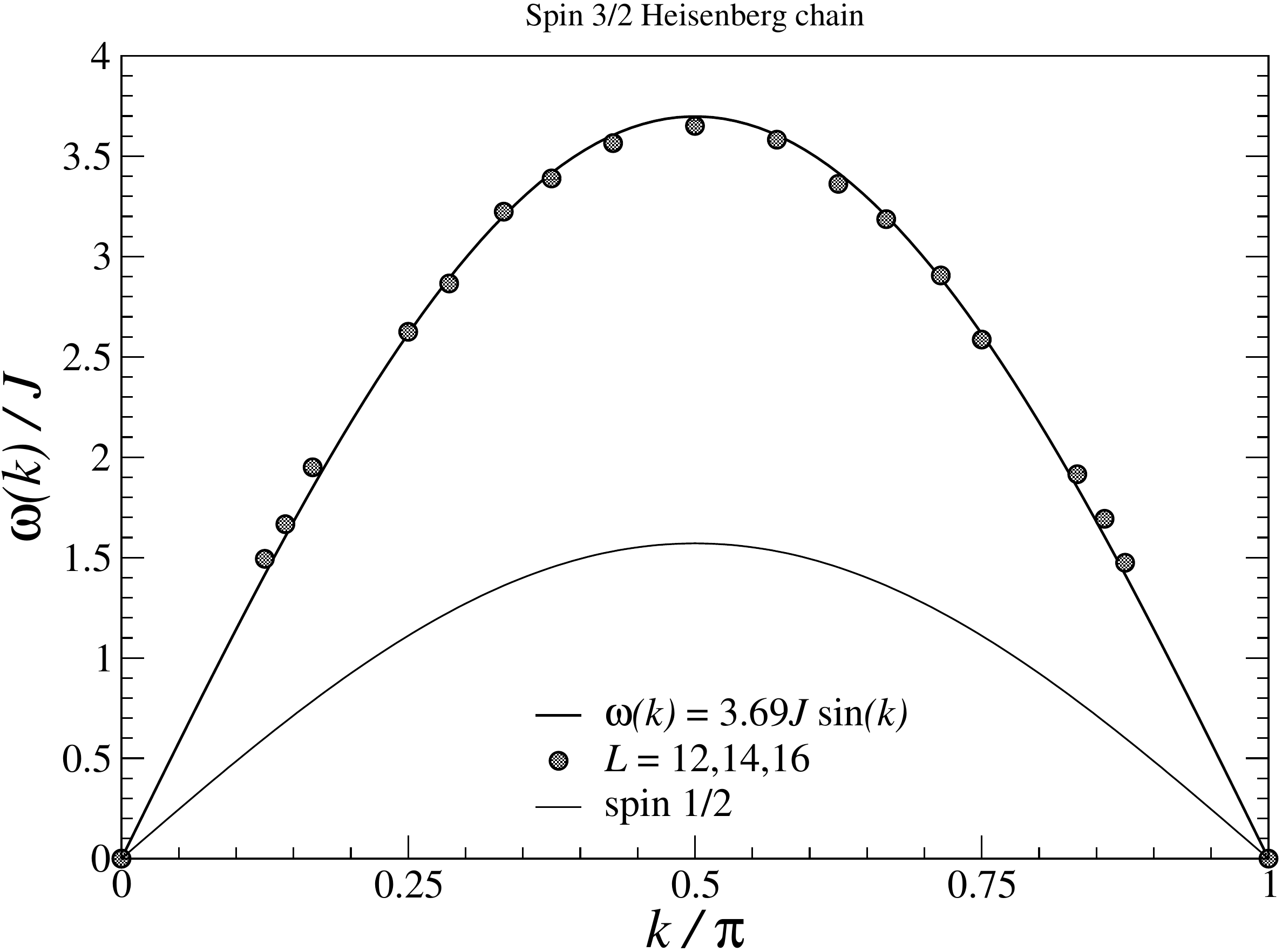}
\caption{Dispersion relation of the spin-3/2 Heisenberg chain.}
\label{fig:omega_k}
\end{figure}

We compute the low-lying excitation spectrum $\omega(k) = E(k)-E_0$ to analyze elementary excitations. The corresponding dispersion relation is shown on Fig.~\ref{fig:omega_k} and shows a very similar structure as the Heisenberg model for spin $1/2$ except that the prefactor is different.
  
\begin{figure}[t]
\centering
\includegraphics[width=0.6\columnwidth,clip]{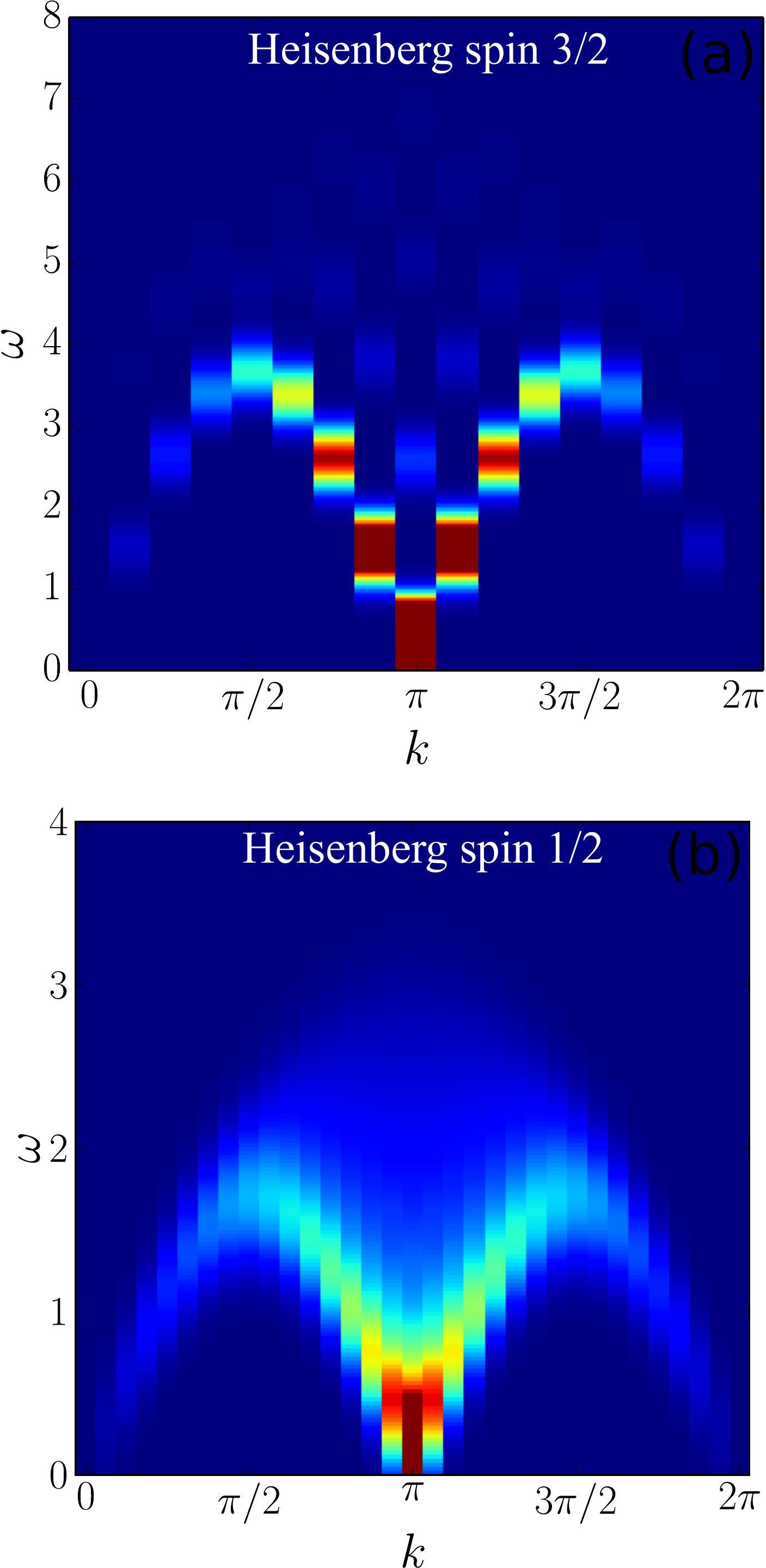}
\caption{Comparison of the dynamical structure factors of the Heisenberg models for (a) the  spin $3/2$ ($L=16$) and (b) spin $1/2$ ($L=32$) chains (J = 1 meV).}
\label{fig:SkwHeisenberg}
\end{figure}

The longitudinal dynamical structure factor at zero-temperature is computed using lorentzian widening of the spectral function
\begin{equation}
\mathcal{S}^{zz}(q,\omega) = \frac 1 L \sum_{n\neq 0} |\elem{n}{S^{z}_q}{0}|^2\delta(\omega-(E_n-E_0))
\end{equation}
where $S^{z}_q = \sum_{r=1}^L e^{iqr} S^{z}_r$ such that $S^{z}_{q=0} = S^{z}_{\text{tot}}$. Results are given in Fig.~\ref{fig:SkwHeisenberg} for the spin $1/2$ and spin $3/2$ Heisenberg models, displaying that the spectral weight of the spin $3/2$ is more concentrated on the magnon arc than for the spin $1/2$ model.
A rough estimate of the leading $J$ coupling from the dispersion relation gives $J \simeq 3.67 \text{meV} \simeq 43$ K.

\begin{figure}[t]
\centering
\includegraphics[width=0.8\columnwidth,clip]{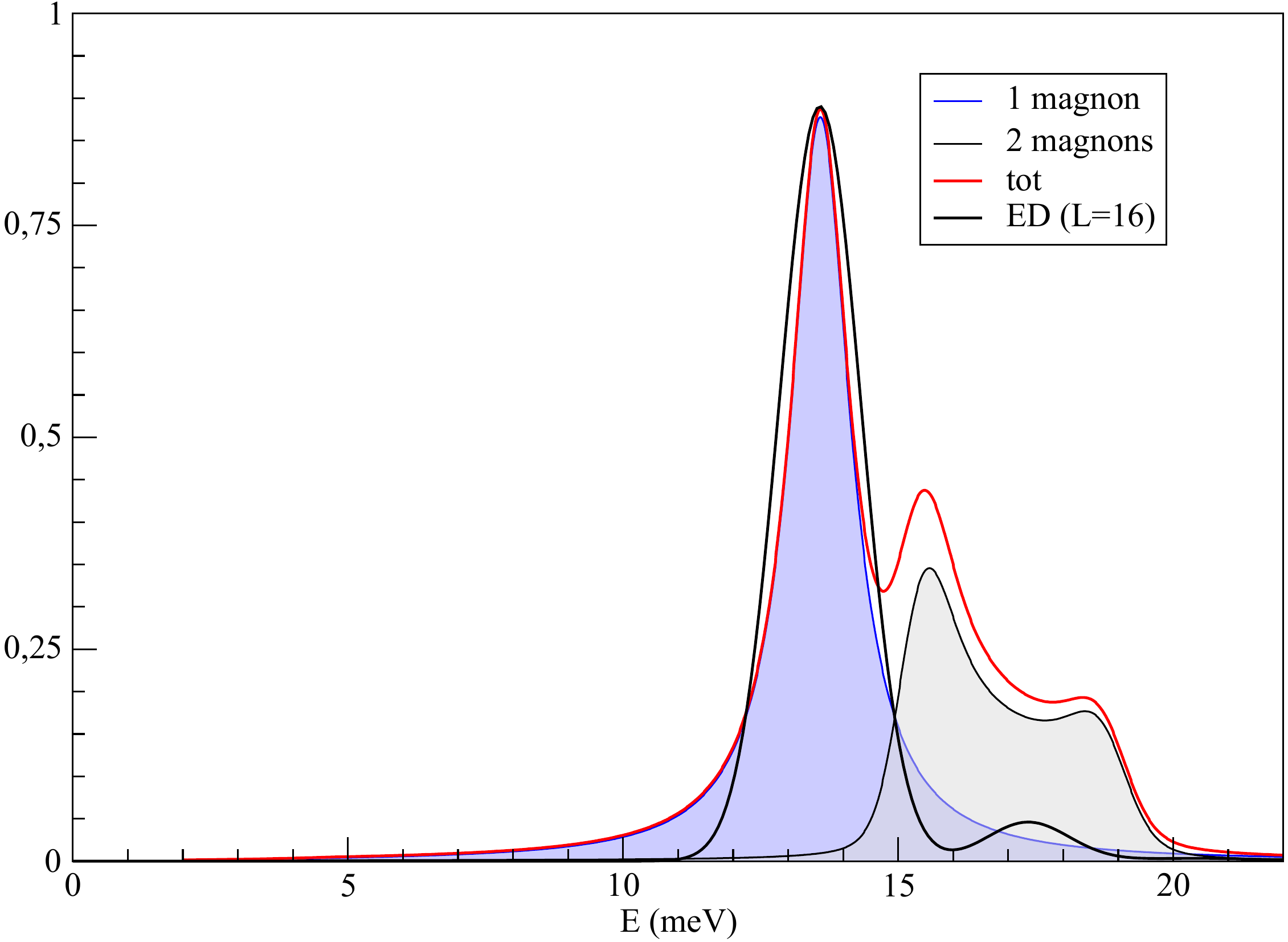}
\caption{Comparing the 1-magnon and 2-magnon modes with ED calculations on $L=16$.}
\label{fig:comparison}
\end{figure}

In Fig.~\ref{fig:comparison}, we compare the multi-magnon model with the ED calculation on a finite system. While the ED does display some weight on multi-magnon processes, it is much less pronounced than the experimentally observed one that is reproduced by the 2 magnon calculation.

\subsection*{Thermodynamics and magnetic susceptibility}

For thermodynamics, we use for the larger sizes propagation a stochastic sampling of the Hilbert space supplemented with imaginary time propagation in each symmetry and $S^{z}_{\text{tot}}$ sectors. We typically use 50 samples of Gaussian random states drawn on the symmetrized basis when the symmetrized Hilbert space size is larger than 500, and compute exactly the thermodynamics for smaller sectors. This easy-to-implement technique allows to capture well the thermodynamics down to fractions of the dominant coupling but fails at low-temperatures.
We use $\beta=1/k_BT$ and for averages $\moy{\cdots}=\textrm{Tr}[(\cdots)e^{-\beta\Ham}/Z]$

For a system with uniaxial anisotropy along $z$, there is both a parallel and transverse susceptibility per site $\chi^{zz}=\chi_{\parallel}(T)$ and $\chi^{xx}=\chi_{\perp}(T)$. For a model with SU(2) symmetry one has $\chi_{\parallel}=\chi_{\perp}$. The zero-field susceptibilities are calculated in the following way
\begin{align}
\chi_{\parallel} &= \frac{\beta}{L}[\moy{(S^z_{\text{tot}})^2}-\moy{S^z_{\text{tot}}}^2] \\
\chi_{\perp} &= \frac{\beta}{L}[\moy{(S^x_{\text{tot}})^2}-\moy{S^x_{\text{tot}}}^2]
\end{align}
and for a $S^z_{\text{tot}}$ conserving Hamiltonian with $S^z_{\text{tot}}\rightarrow -S^z_{\text{tot}}$ symmetry, we have $\moy{S^z_{\text{tot}}}=\moy{S^x_{\text{tot}}}=0$.


With our definitions, the large-$T$ tail of the susceptibility directly gives access to the spin of the model since, for an isotropic model,
\begin{align}
\chi^{\alpha\alpha} \simeq \frac{S(S+1)}{3k_BT}
\end{align}

For the magnetic susceptibility of the previously studied powder compound, we use the following relation to estimate the coupling $J$ from experimental data
\begin{align}
\chi_{\text{exp}}(T) = \frac{\mu_0(g\mu_B)^2}{k_B}\frac{k_B}{J} \chi_{\text{th}}\left(\frac{k_BT}{J}\right)
\end{align}
in which the coupling $J/k_B$ will be set in Kelvin and is a free parameter.
A rough estimate based on $L=8$ numerics and the maximum position of $\chi$ gives $J \simeq  45 \text{K} \simeq 3.87$ meV which is compatible with the neutron scattering data.

\subsection*{XXZ model}

The dynamical structure factor was computed using exact diagonalisation for a S = 3/2 chain system, considering an XXZ-type anisotropy in the following Hamiltonian: 

\begin{equation}
H= J \sum_{i,j} \left[  \vec{S}_{i}\cdot \vec{S}_{j} + \delta_{z} S_{i}^{z}S_{j}^{z} \right].
\label{eq:alpha}
\nonumber
\end{equation}

\begin{figure}[t]
\centering
\includegraphics[width=1\columnwidth,clip]{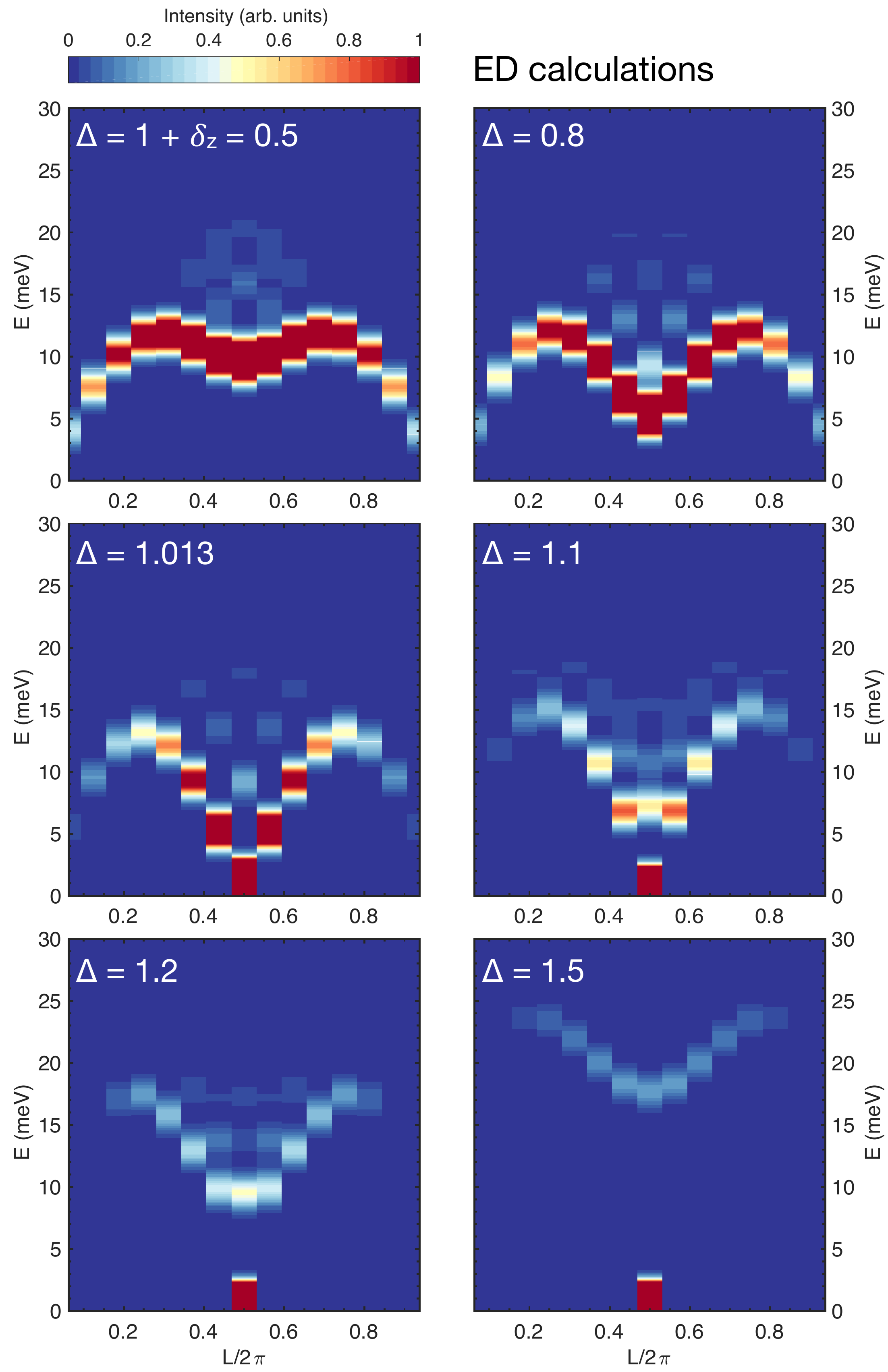}
\caption{Dynamical structure factor using the XXZ model as a function of $\Delta = 1+\delta_z$.}
\label{fig:EDvsdelta}
\end{figure}

Several values for $\Delta = 1 + \delta_z$ were tested with both $\delta_z$ positive (Ising-like anisotropy) and negative (planar anisotropy), and the results are displayed in Figure \ref{fig:EDvsdelta}. A value of $\Delta$ significantly different from the Heisenberg limit strongly affects the distribution of spectral weight as well as the presence of an energy gap at $L = \pi$. It is worth noting that the ED calculations account for both single and multi-magnon processes and the intensity is therefore distributed between the different processes. Identifying the spectral weight associated to the single and two-magnon scattering would however require a further understanding of the microscopic parameters in this system. $\Delta = 1.013$ corresponds to the value extracted from the spin-wave analysis presented above and the ED result shows that such small anisotropy does not significantly change the spectrum, compared to the Heisenberg isotropic model. One must keep in mind that the ED spectra are naturally discrete and although the spectral weight distribution of intensity carry information close to the thermodynamic limit, a continuum of intensity is generally expected. The discrete nature of the spectra is therefore reminiscent of the finite-size of the system, limiting the calculations.

\bibliography{CaCrO}